\documentclass[fleqn,usenatbib]{mnras}
\usepackage{lineno}

\usepackage{newtxtext,newtxmath}

\usepackage[T1]{fontenc}
\usepackage{fix-cm}
\usepackage{xcolor}

\DeclareRobustCommand{\VAN}[3]{#2}
\let\VANthebibliography\thebibliography
\def\thebibliography{\DeclareRobustCommand{\VAN}[3]{##3}\VANthebibliography}

\usepackage{graphicx}	
\usepackage{amsmath}	
\usepackage{booktabs}
\usepackage{multirow}
\usepackage{hyperref}

\title[The Evolution of X-ray Spectra in TDEs]{The Evolution of X-ray Spectra in Tidal Disruption Events}

\author[Chen et al.]{
Wei Chen$^{1,2}$
and Erlin Qiao$^{1,2}$
\thanks{E-mail: qiaoel@nao.cas.cn}
\\
$^{1}$National Astronomical Observatories, Chinese Academy of Sciences, Beijing 100101, China\\
$^{2}$School of Astronomy and Space Sciences, University of Chinese Academy of Sciences, 19A Yuquan Road, Beijing 100049, China 
}

\date{Accepted XXX. Received YYY; in original form ZZZ}

\pubyear{\the\year{}}

\begin{document}
\label{firstpage}
\pagerange{\pageref{firstpage}--\pageref{lastpage}}
\maketitle

\begin{abstract}
The study of the evolution of X-ray spectra in tidal disruption events (TDEs) is an important approach for understanding the physical processes occurring near a supermassive black hole. Observations show that the X-ray spectra of TDEs are very soft at the peak after the outburst, followed by a spectral hardening on a timescale of years. Theoretically, TDEs are suggested to undergo super-Eddington accretion at the time around the outburst. In this paper, we constructed a new disc-corona model to explain the observed X-ray spectral hardening in TDEs. In our model, there is a transition radius $r_{\text{tr}}$. For $r< r_{\text{tr}}$, the accretion flow exists in the form of a slim disc, the emission of which is dominated by soft X-rays. While for $r>r_{\text{tr}}$, the accretion flow exists in the form of a traditionally sandwiched disc-corona, in which a harder X-ray spectrum is produced. Our calculations show that $r_{\text{tr}}$ decreases with decreasing mass accretion rate $\dot {M}$, which intrinsically can predict the hardening of the X-ray spectra since the relative contribution of the outer disc-corona to the inner slim disc to the X-ray spectrum increases with decreasing $\dot {M}$. Our model has been applied to explain the observed X-ray spectral hardening in  TDE candidate AT 2019azh, in which $\dot {M}$ is assumed to decrease proportionally to $t^{-5/3}$. Potential applications of the model in explaining the X-ray spectral evolution in upcoming rich TDE observations are also expected.
\end{abstract}




\section{Introduction} \label{sec:intro}
When a star moves close to a massive black hole, it will be disrupted if its self-gravity is insufficient to counteract the tidal forces exerted by the black hole (BH) \citep{1975Natur.254..295H}. Generally, approximately half of the stellar material is ejected, while the remaining half is bound to the BH. This bound debris is gradually circularized and accreted onto the BH.

TDEs were originally discovered in the X-ray band, and X-ray emission is generally believed to be powered by accretion \citep{,1999ApJ...514..180U,2002RvMA...15...27K}. The X-ray light curves exhibit a rapid rise to peak luminosities of 
$10^{42} \sim 10^{44} \, \text{erg} \, \text{s}^{-1}$, followed by a gradual decline of `$t^{-5/3}$' as the predicted fallback rate \citep{ 1989IAUS..136..543P,1990Sci...247..817R, 2011MNRAS.410..359L}.
Near the peak, the X-ray spectra are very soft, with blackbody temperatures in the range of $kT_{\text{bb}} = 0.04 \sim 0.12 \, \text{keV}$, or power-law indices ranging from $\Gamma = 3 \sim 5$ \citep{2015JHEAp...7..148K,2021SSRv..217...18S}. Over time, the spectra harden: for instance, RXJ 1242-1119 shows a spectral evolution from $\Gamma \sim 5$ to $\Gamma \sim 2.5$ over several years \citep{2004ApJ...603L..17K}. A similar hardening from $\Gamma \sim 4.0$ to $\Gamma \sim 2.4$ occurred in NGC 5905 three years after the peak \citep{1999A&A...343..775K}, and RBS 1032 exhibited a spectral hardening from $\Gamma \sim 5$ to $\Gamma \sim 3.4$ over two decades \citep{2014ApJ...792L..29M}. The variation in the spectral index suggests that the accretion rate is evolving over time. Theoretically, TDEs are suggested to undergo a phase of intense super-Eddington accretion in their early stages \citep{1988Natur.333..523R,2009MNRAS.400.2070S}, where the Eddington accretion rate is defined as $\dot{M}_{\text{Edd}} = L_{\text{Edd}} / (0.1 c^2) = 1.39 \times 10^{18} \, M/M_{\odot} \, \text{g} \, \text{s}^{-1}$ with the Eddington luminosity given by $L_{\text{Edd}} = 1.25 \times 10^{38} \left(M/M_{\odot} \right) \, \text{erg} \, \text{s}^{-1}$ (where $M$ is the BH mass). At super-Eddington rates ($\dot{M} / \dot{M}_{\text{Edd}} \geq 1$), the accretion flow is expected to exist in the form of a geometrically and optically thick slim disc \citep{1988ApJ...332..646A}.
As time progresses, the accretion rate decreases to sub-Eddington values ($0.01 \leq \dot{M}_{\text{Edd}} \leq 1$), and the disc transitions to a geometrically thin, optically thick standard disc \citep{1973A&A....24..337S}.

However, the physical cause of the TDE spectral hardening is still under debate. Some studies suggest that this is due to the formation of a corona \citep{2021MNRAS.504.4730M, 2021ApJ...912..151W, 2022ApJ...937....8Y, 2023MNRAS.519.2375C, 2024ApJ...970..116W,2024ApJ...966..160G}. In the traditional disc-corona model, hard X-ray emission is attributed to a hot corona located above a cool standard accretion disc, which has been used to explain the strong hard X-ray emission observed in luminous active galactic nuclei (AGNs) and X-ray binaries \citep{1991ApJ...380L..51H, 1993ApJ...413..507H, 1994ApJ...436..599S, 1996ApJ...470..249P, 2002ApJ...572L.173L, 2003ApJ...587..571L,2012ApJ...754...81L,2012ApJ...759...65T,2013ApJ...777..102Q,2017MNRAS.467..898Q,2018MNRAS.477..210Q,2013ApJ...764....2Q}. Hard X-ray emission follows a power-law and is thought to be generated by inverse Compton scattering of soft photons from the accretion disc in a hot corona above \citep{1994ApJ...436..599S,1998MNRAS.301..179M,2000ApJ...542..703Z,2016MNRAS.458.2454L}. One of the most promising mechanisms for coronal heating is magnetic reconnection, which arises from the dynamo-generated magnetic fields in the accretion disc and is facilitated by Parker buoyancy instability, allowing stored magnetic energy to be released in the corona, ultimately radiated as X-rays \citep{1992MNRAS.259..604T, 1998MNRAS.299L..15D, 2000ApJ...534..398M, 2002ApJ...572L.173L, 2003ApJ...587..571L, 2009MNRAS.394..207C, 2012ApJ...761..109Y}. However, the formation of the corona during super-Eddington accretion in TDEs remains an open question for investigation.

This work aims to explain the observed X-ray emission and the spectral hardening in TDEs by developing a new disc-corona model within the context of super-Eddington accretion. Building upon the existing magnetic-reconnection-heated corona model \citep{2002ApJ...572L.173L, 2003ApJ...587..571L, 2016ApJ...833...35L, 2020MNRAS.495.1158C},
we replace the standard disc in the inner region with a slim disc, providing a more accurate description of the TDE environment. In our model, the accretion flow in the inner region is dominated by soft X-ray emission from the slim disc, while in the outer region, the accretion flow is characterized by a traditional disc-corona configuration, which produces a harder X-ray spectrum. Our calculations show that as $\dot {M}$ decreases, X-ray spectral hardening naturally occurs. This is because the relative contribution of the outer disc-corona to the overall X-ray spectrum increases as $\dot {M}$ decreases, while the soft X-ray emission from the inner slim disc weakens. We compare our theoretical predictions with the observed X-ray data in TDE candidate AT 2019azh, finding good agreement with the observed spectral evolution.

This paper is organized as follows.  
We introduce the new disc-corona model in Section \ref{sec:2}.  
In Section \ref{sec:3}, we present the results, focusing on the evolution of the emergent spectra with decreasing $\dot{M}$ in the environment of TDEs, and compare the theoretical predictions with the observed X-ray data for TDE candidate AT 2019azh. Finally, the summary and discussion are presented in Section \ref{sec:4}.

\section{The Model} \label{sec:2}
\subsection{A new Disc-corona Model}\label{sec:2.1}

In this paper, based on the magnetic-reconnection-coupled disc-corona model in \citet{2002ApJ...572L.173L, 2003ApJ...587..571L}, we proposed a new disc-corona model. We will first give a brief summary of the disc-corona model in \citet{2002ApJ...572L.173L, 2003ApJ...587..571L}.
In \citet{2002ApJ...572L.173L, 2003ApJ...587..571L}, the standard accretion disc is coupled with the corona by considering both mass exchange and radiative coupling between the disc and the corona. In the specific calculations, the disc is assumed to be either gas pressure-dominated or radiation pressure-dominated separately for simplicity.
In the disc, magnetic fields are continuously generated via dynamo action. Further, the magnetic flux loops emerge into the corona and reconnect with other loops, resulting in the corona being heated. The heat in the corona will be conducted downward by electrons to the chromosphere above the disc, causing the matter in the disc to evaporate into the corona.
With the increase of the coronal density, the cooling of the corona is dominated by inverse Compton scattering of the soft photons from the accretion disc in the corona.
Eventually, an equilibrium between the heating and the cooling to the corona is reached, and a stable disc-corona structure is established. The equations describing these processes in the corona and at the
interface of disc and corona are,
\begin{equation}
\label{Equ:equ1}
\frac{B^{2}}{4\pi}V_{\text{A}}\approx \frac{4kT}{m_{\text{e}}c^{2}}\tau^{\ast} c U_{\text{rad}},
\end{equation}
\begin{equation}
\label{Equ:equ2}
\frac{k_{0}T^{7/2}}{l_{\rm c}}\approx \frac{\gamma}{\gamma - 1}nkT(\frac{kT}{m_{\text{H}}})^{1/2},
\end{equation}
Equation (\ref{Equ:equ1}) describes the energy balance in the corona between magnetic heating and Compton cooling.  
The magnetic loops emerge at the Alfvén speed $V_{\text{A}}=\sqrt{B^2/4\pi\mu m_{\text{H}} n}$, carrying magnetic energy flux $\frac{B^{2}}{4\pi}V_{\rm A}$ from the disc into the corona, which is ultimately radiated away by inverse Compton scattering. Here, $B$ is the magnetic field strength, $c$ is the speed of light, $T$, $n$ and $\tau^*$ are the temperature, number density and effective optical depth of the corona. $U_{\rm rad}$ is the energy density of soft photons. Equation (\ref{Equ:equ2}) describes the energy balance between conductive flux by electrons and enthalpy flux in the chromospheric layer, which can be rewritten as the physical quantities of the corona assuming $n_{\rm evap}h = n l_{\rm c}$ \citep[]{2002ApJ...572L.173L}. Here, $n_{\rm evap}$ is the number density of evaporated matter in the chromosphere, $h$ is the thickness of the chromosphere layer, $l_{\rm c}$ is the length of the magnetic loop in the corona. We take $l_{\rm c} = 10 R_{\text{S}}$ (where $R_{\text{S}} = 2GM/c^{2}$ is the Schwarzschild radius). These two equations together determine the coronal temperature and density for given magnetic field strength and soft photon energy density.

The constants in the above equations are the Boltzmann constant $k = 1.38\times 10^{-16}\text{erg} \ \text{K}^{-1} $, the thermal conductivity coefficient $k_0 = 10^{-6}\text{erg} \ \text{cm}^{-1}\ \text{K}^{-7/2}$, the adiabatic index $\gamma = 5/3$ and the molecular weight for pure hydrogen plasma $\mu =0.5$. The effective optical depth $\tau^*=\lambda_{\tau}\tau=\lambda_{\tau}n\sigma_{\text{T}}l_{\rm c}$ is introduced to account for the isotropic incident photons undergoing upscattering in a sandwich corona geometry and the multiple scattering of soft photons, the correction factor $\lambda_{\tau}$ is therefore a geometrical factor with a value around unity.

We assume that the gas pressure and magnetic pressure in the disc are equally distributed,
\begin{equation}
\label{Equ:equ3}
\beta = \frac{n_{\rm d}kT_{\rm d}}{B^{2}/8\pi} \sim 1
\end{equation}
where $n_{\rm d}$ and $T_{\rm d}$ are the mid-plane number density and temperature of the accretion disc, with which the magnetic field $B$ in the accretion disc can be calculated accordingly.

In this paper, we construct a new disc–corona model by replacing the standard accretion disc with a slim disc if the disc is dominated by radiation pressure.
Specifically, we take the self-similar solution of the slim disc for the gas temperature $T_{\rm d}$ and the number density $n_{\rm d}$ in the disc \citep{2006ApJ...648..523W} (one can also refer to equations (\ref{Equ:equA11}) and (\ref{Equ:equA12}) for details), based on which we derive
the magnetic field as,
\begin{equation}
\label{Equ:equ4}
\begin{split}
B_{1} & = 1.30 \times 10^{2} f_{\text{a}}^{-13/16} \alpha_{0.1}^{-5/8} \beta_{1}^{-1/2} m_{6}^{-5/8}\\
& \times \dot m^{5/8} (1-f_{\text{c}})^{-1} \hat{r}_{10}^{-17/16} \phi^{-3/16} \text{G}
\end{split}
\end{equation}
where $m_{6}=M / (10^{6}M_{\odot})$, $\hat r_{10} = R / (10R_{\text{S}})$, $\dot m = \dot M /\dot M_{\text{Edd}}$, $\phi = 1 - (3R_{\text{S}}/R)^{1/2}$, $\alpha_{0.1}$, $\beta_{1}$ are in units of 0.1, 1, and $f_{\text{a}}$ is defined as equation (\ref{Equ:equA9}). 
For the regions of the accretion disc dominated by gas pressure, we retain the results derived from the standard disc calculations by \citet{2003ApJ...587..571L},
\begin{equation}
\label{Equ:equ5}
\begin{split}
B_{2} & = 1.43 \times 10^{6} \alpha_{0.1}^{-9/20} \beta_{1}^{-1/2} m_{6}^{-9/20}\\
& \times [\dot m (1-f_{\text{c}})]^{2/5} \hat{r}_{10}^{-51/40} \phi^{2/5} \text{G} 
\end{split}
\end{equation}
Here, $f_{\text{c}}$ is the energy fraction dissipated in the corona. In the regions of the accretion disc dominated by radiation pressure, we define,
\begin{equation}
\label{Equ:equ6}
f_{\text{c}} = \frac{F_{\text{cor}}}{F_{\text{rad}}} = (\frac{B^2}{4\pi}V_{\text{A}})(\sigma T_{\text{eff}}^{4})^{-1}
\end{equation}
where $F_{\text{cor}}$ is the accretion energy released in the reconnected magnetic corona, $F_{\text{rad}}$ is the radiation flux. In the regions of the accretion disc dominated by gas pressure, as in \citet{2003ApJ...587..571L} we define,
\begin{equation}
\label{Equ:equ7}
f_{\text{c}} = \frac{F_{\text{cor}}}{F_{\text{tot}}} = (\frac{B^2}{4\pi}V_{\text{A}})(\frac{3GM\dot M \phi}{8\pi R^{3}}
)^{-1}
\end{equation}

In Equation (\ref{Equ:equ1}), the energy density of soft photons $U_{\text{rad}}$ consists of two components: intrinsic disc radiation $U_{\text{rad}}^{\text{in}}$ and reprocessed radiation $U_{\text{rad}}^{\text{re}}$ of downward Compton,
\begin{equation}
\label{Equ:equ8}
U_{\rm rad}^{\rm in} = a T_{\rm eff}^{4}= 7.93 \times 10^{8} f_{\rm a}^{1/2} (1-f_{\rm c})^{-1} m_6^{-1} \hat{r}{10}^{-2} \phi^{1/4} \ \text{erg cm}^{-3}
\end{equation}
\begin{equation}
\label{Equ:equ9}
U_{\text{rad}}^{\text{re}}=0.4\lambda_{\text{u}}\frac{B^{2}}{8\pi}
\end{equation}
$\lambda_{\text{u}}$ is a factor introduced for the evaluation of the seed field in \citet{1991ApJ...380L..51H, 1993ApJ...413..507H}, it accounts for deviations from isotropic scattering and the difference between the speed of magnetic energy release (at the Alfvén speed) and the speed of radiative transport (at the speed of light). The value of which is around unit in order of magnitude. And $T_{\text{eff}}$ is determined by equation (\ref{Equ:equA14}). By solving equations (\ref{Equ:equ1}), (\ref{Equ:equ2}), (\ref{Equ:equ4}), and (\ref{Equ:equ8}), we derive the solution for the corona above the radiation pressure-dominated disc,
\begin{equation}
\begin{split}
\label{Equ:equ10}
T_{1}=1.36 & \times 10^{7}f_{\text{a}}^{-47/64}\alpha_{0.1}^{-15/32}\beta_{1}^{-3/8}m_{6}^{-3/32}\dot m ^{15/32} \\
& \times (1-f_{\text{c}})^{-1}\hat{r}_{10}^{-19/64}\lambda_{\tau}^{-1/4}l_{c,10}^{1/8}\phi^{-17/64}\text{K}
\end{split}
\end{equation}
\begin{equation}
\begin{split}
\label{Equ:equ11}
n_{1}=1.99 & \times 10^{7}f_{\text{a}}^{-47/32}\alpha_{0.1}^{-15/16}\beta_{1}^{-3/4}m_{6}^{-19/16}\dot m ^{15/16} \\
& \times (1-f_{\text{c}})^{-2}\hat{r}_{10}^{-19/32}\lambda_{\tau}^{-1/2}l_{c,10}^{-3/4}\phi^{-17/32}\text{cm}^{-3}
\end{split}
\end{equation}
where $l_{\rm c}$ is in units of $10R_{\text{S}}$. By solving equations (\ref{Equ:equ1}), (\ref{Equ:equ2}), (\ref{Equ:equ5}), and (\ref{Equ:equ9}), we derive the solution for the corona above the gas pressure-dominated disc,
\begin{equation}
\begin{split}
\label{Equ:equ12}
T_{2}=5.76 & \times 10^{9}\alpha_{0.1}^{-9/80}\beta_{1}^{-1/8}m_{6}^{1/80}\dot m ^{1/10} \\
& \times (1-f_{\text{c}})^{1/10}\hat{r}_{10}^{-51/160}\lambda_{\tau}^{-1/4}\lambda_{\text{u}}^{-1/4}l_{c,10}^{1/8}\phi^{1/10}\text{K}
\end{split}
\end{equation}
\begin{equation}
\label{Equ:equ13}
\begin{split}
n_{2}=3.59 & \times 10^{12}\alpha_{0.1}^{-9/40}\beta_{1}^{-1/4}m_{6}^{-39/40}\dot m ^{1/5} \\
& \times (1-f_{\text{c}})^{1/5}\hat{r}_{10}^{-51/80}\lambda_{\tau}^{-1/2}\lambda_{\text{u}}^{-1/4}l_{c,10}^{-3/4}\phi^{1/5}\text{cm}^{-3}
\end{split}
\end{equation}

By substituting equations (\ref{Equ:equ4}) and (\ref{Equ:equ11}) into equation (\ref{Equ:equ6}), we obtain an equation for the coronal energy fraction $f_{\text{c}}$ in the radiation pressure-dominated region of the accretion disc,
\begin{equation}
\label{Equ:equ14}
\begin{split}
f_{\text{c}}=1.44 & \times 10^{-6}f_{\text a}^{-141/64}\alpha_{0.1}^{-45/32}\beta_{1}^{-9/8}m_{6}^{-9/32}\dot m ^{45/32} \\
& \times (1-f_{\text{c}})^{-2}\hat{r}_{10}^{-57/64}\lambda_{\tau}^{1/4}l_{c,10}^{3/8}\phi^{-83/64} \\
& = c_{1}(1-f_{\text{c}})^{-2}
\end{split}
\end{equation}
By substituting equations (\ref{Equ:equ5}) and (\ref{Equ:equ13}) into equation (\ref{Equ:equ7}), we obtain an equation for the coronal energy fraction $f_{\rm c}$ in the gas pressure-dominated region $f_{\text{c}}$ of the accretion disc as in \citet{2003ApJ...587..571L},
\begin{equation}
\label{Equ:equ15}
\begin{split}
f_{\text{c}}=3.15 & \times 10^{4}\alpha_{0.1}^{-99/80}\beta_{1}^{-11/8}m_{6}^{11/80}\dot m ^{1/10} \\
& \times (1-f_{\text{c}})^{11/10}\hat{r}_{10}^{-81/160}\lambda_{\tau}^{1/4}\lambda_{\text{u}}^{1/4}l_{c,10}^{3/8}\phi^{1/10}
\end{split}
\end{equation}
Noting that equation~(\ref{Equ:equ15}) has solutions for any accretion rates, while equation~(\ref{Equ:equ14}) has solutions only under the condition $c_1 \leq 4/27$.

Given the values of $M$, $\dot{M}$, $R$, $\alpha$, $\beta$, initial parameters $\lambda_{\tau}=1$ and $\lambda_{\text{u}}=1$, we solve equation (\ref{Equ:equ14}) (or eq. (\ref{Equ:equ15})) for $f_{\text{c}}$, and determine the coronal quantities from equation (\ref{Equ:equ10}) and (\ref{Equ:equ11}) (or eqs. (\ref{Equ:equ12}) and (\ref{Equ:equ13})).

\subsection{The spectra calculated from Monte Carlo Simulations}
Based on the parameters of the disc–corona structure derived in Section \ref{sec:2.1}, we can utilise the Monte Carlo method to calculate the emergent spectrum of the disc–corona system \citep{1977SvA....21..708P,1983ASPRv...2..189P}. The blackbody radiation emitted by the accretion disc at a radius $R$ with a temperature $T_{R}$ is given by,
\begin{equation}
\label{Equ:equ16}
\begin{split}
\sigma T_{R}^{4}=\frac{3GM \dot M \phi (1-f_{\text{c}})}{8 \pi R^{3}}+\frac{c}{4}U_{\text{rad}}^{\text{re}}\\
\approx \text{max}[\frac{3GM \dot M \phi (1-f_{\text{c}})}{8 \pi R^{3}},\frac{c}{4}U_{\text{rad}}^{\text{re}}]
\end{split}
\end{equation}

To obtain a self-consistent solution for the disc–corona model, in the case where the accretion disc is dominated by gas pressure, we need to check the self-consistency in two aspects: (1) the total energy of the photons emitted downward from below the corona (toward the accretion disc) should be approximately equal to the total energy of soft photons in structural calculations, i.e., $L_{\text{down}} \approx L_{\text{soft}}$, where $L_{\text{soft}}=\int_{R_{\text{in}}}^{R_{\text{out}}}2 \pi R \sigma T_{R}^4 dR$. And (2) the total energy of the photons emitted upward from above the corona (as observed by the observer) should be approximately equal to the total gravitational energy released, i.e., $L_{\text{up}} \approx L_{\text{G}}$, where $L_{\text{G}}=\int_{R_{\text{in}}}^{R_{\text{out}}}2 \pi R (3GM \dot M \phi/8 \pi R^{3}) dR$. If the consistency condition is satisfied, we consider that we have found suitable values for $\lambda_{\text{u}}$ and $\lambda_{\tau}$. If not satisfied, we update the values as follows: $\lambda_{\text{u},n+1} = \left(\frac{L_{\text{down},n}}{L_{\text{soft},n}}\right) \lambda_{\text{u},n}$ and $\lambda_{\tau,n+1} = \left(\frac{L_{\text{up},n}}{L_{\text{soft},n}}\right) \lambda_{\tau,n}$. We then repeat the numerical calculations and Monte Carlo simulations until the consistency conditions $L_{\text{up}} \approx L_{\text{G}}$ and $L_{\text{down}} \approx L_{\text{soft}}$ are satisfied. In this case, the converged values of $\lambda_{\text{u}}$ are typically in the range of $1.0\sim6.0$ as in Table 1 of \citet{2003ApJ...587..571L}, increasing with accretion rate, while $\lambda_{\tau}$ is approximately 1.4 and nearly independent of accretion rate. These values reflect geometrical effects and multiple Compton scatterings in the slab corona.

In the case where the accretion disc is dominated by radiation pressure, we just need to ensure that the total energy of the photons emitted upward from above the corona (as observed by the observer) is approximately equal to the radiated energy, i.e., $L_{\text{up}} \approx L_{\text{rad}}$ where $L_{\text{rad}} = \int_{R_{\text{in}}}^{R_{\text{out}}}2\pi R F_{\text{rad}} dR=\int_{R_{\text{in}}}^{R_{\text{out}}}2 \pi R \sigma T_{\text{eff}}^{4} dR$. We only need to iteratively solve for $\lambda_{\tau}$, while $\lambda_{\text{u}}$ can be set to unity \citep{2002ApJ...572L.173L, 2003ApJ...587..571L}. In this case, the Compton scattering in the
corona is quite weak, and we find $\lambda_{\tau} \sim 1$.

\begin{figure*}
\centering
\includegraphics[scale=0.47]{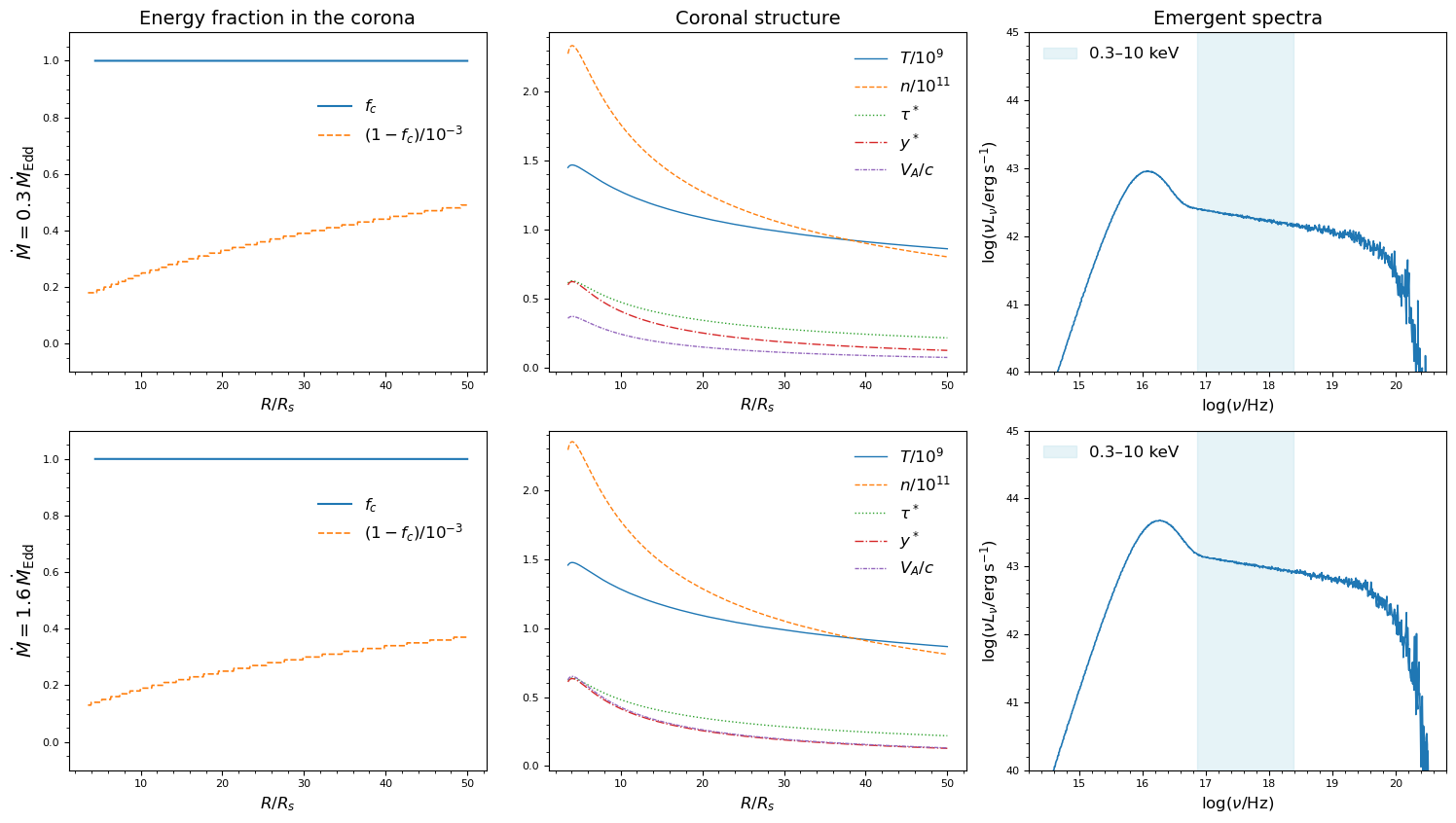}
\caption{Gas pressure-dominated solution for $\dot{m}=0.3$ (top row) and $\dot{m}=1.6$ (bottom row). In this calculation, we fix $M=10^{6}M_{\odot}$, $\alpha=0.3$, $\beta=1$. \textbf{Left panels:} Fraction of energy dissipated in the corona ($f_{\rm c}$) and in the disc ($1-f_{\rm c}$). \textbf{Middle panels:} Radial profiles of temperature ($T/10^9$ K), number density ($n/10^{11}$ cm$^{-3}$), optical depth ($\tau$), Compton $y$-parameter, and Alfv\'en speed ($V_A/c$). \textbf{Right panels:} Emergent spectra corresponding to the two accretion rates. 
The light blue shading marks the 0.3-10 keV band.}  
\label{Fig:Fig1}
\end{figure*}

\subsection{The self-consistent solutions} \label{sec:2.3}
In the following, we show the results of the basic solutions of our new disc-corona model, i.e. the gas pressure-dominated solution and radiation pressure-dominated solution respectively.
In the calculations, we fix $\alpha=0.3$, $\beta=1$, the inner radius of the accretion disc $R_{\rm in}=3R_{\rm S}$ for a non-rotating BH, and the outer radius of the accretion disc $R_{\rm out}=50R_{\rm S}$.

Gas pressure-dominated solution:
This solution is the same as that shown in \citet{2003ApJ...587..571L}, in which the accretion energy is nearly completely dissipated in the corona, i.e. $f_{\rm c}\sim 1$, and the X-ray spectra are relatively hard with a hard X-ray (2-10 keV) photon index $\Gamma \sim 2.1$. One can refer to Fig. 1 and Fig. 5 in \citet{2003ApJ...587..571L} for details. There are two key conclusions for the gas pressure-dominated solution, i.e., (1) gas pressure-dominated solution can exist for any $\dot m$; (2) X-ray spectra nearly do not change with increasing $\dot m$ with a hard X-ray photon index $\Gamma \sim 2.1$.
Here are examples, in Fig. \ref{Fig:Fig1} we show the fraction of the energy dissipated in the corona $f_{\rm c}$ (left panel), the coronal structure (middle panel) (including electron temperature $T/10^{9}$, number density $n/10^{11}$, effective optical depth $\tau^{}$, Compton $y$ parameter ($y^{}=4kT/m_{e}c^{2}$) and Alfv\'en speed $V_A/c$ in the corona) and emergent spectra (right panel) for $\dot{m}=0.3$ and $\dot m=1.6$ by fixing $M=10^{6}M_{\odot}$.

\begin{figure}
\centering
\includegraphics[scale=0.49]{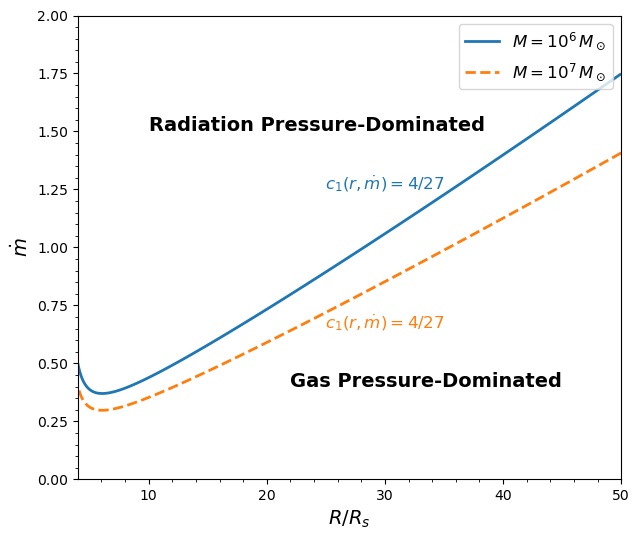}
\caption{Mass accretion rate $\dot m$ as a function of radius $R/R_{\rm S}$ by setting $c_1 = 4/27$ for $M=10^{6}M_{\odot}$ (blue solid line) and $M=10^{7}M_{\odot}$ (orange dashed line) respectively, i.e. the critical (minimum) $\dot m$ for the existence of the radiation pressure-dominated solution at a fixed radius for $M=10^{6}M_{\odot}$ and $M=10^{7}M_{\odot}$ respectively.
The region above the curve of $\dot m$ as a function of $R/R_{\rm S}$
corresponds to the radiation pressure-dominated solution, and the region below the curve of $\dot m$ as a function of $R/R_{\rm S}$ corresponds to the gas pressure-dominated solution.}
\label{Fig:fig2}
\end{figure}

\begin{figure*}
\centering
\includegraphics[scale=0.47]{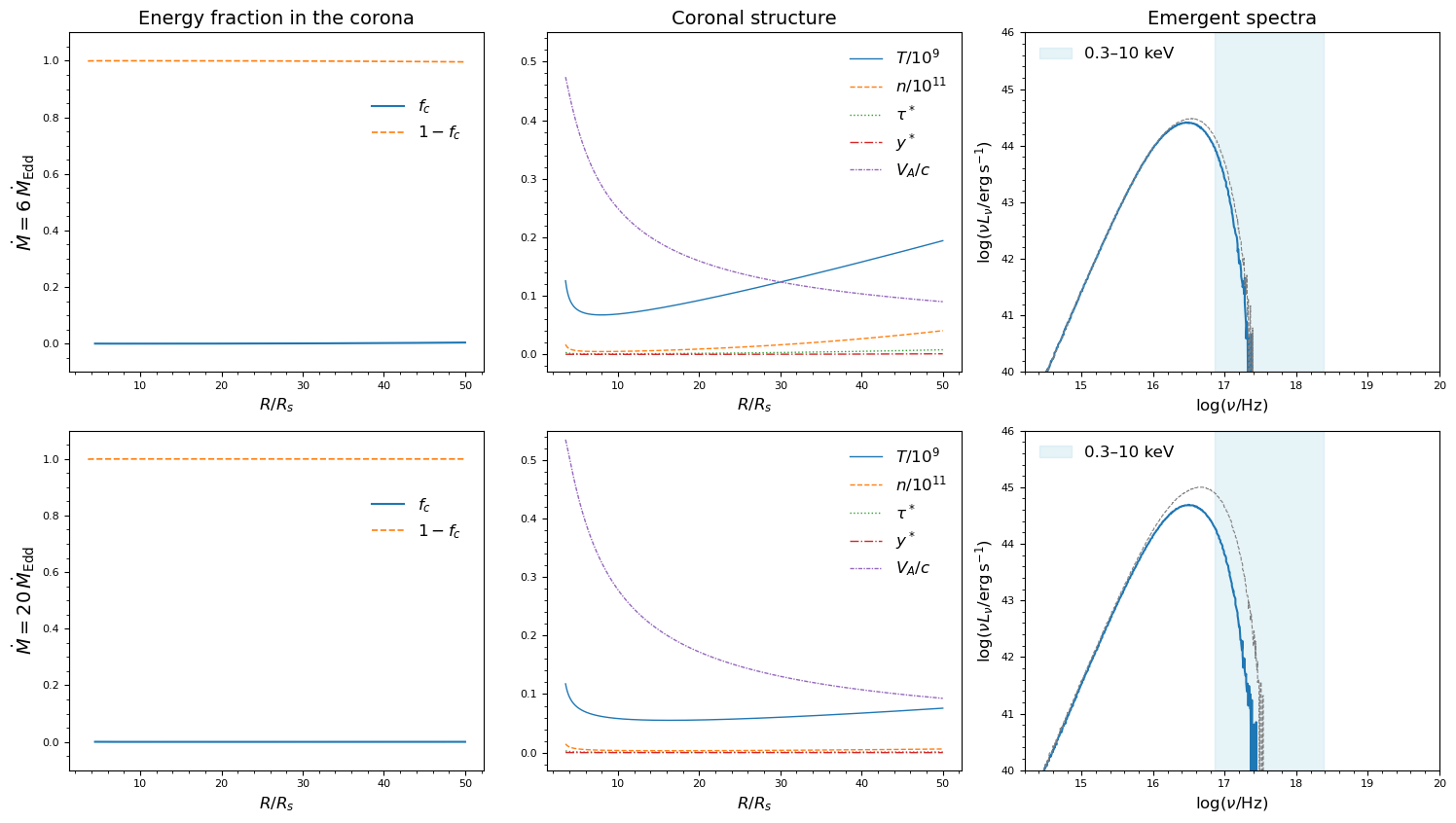}
\caption{Radiation pressure-dominated solution for $\dot{m}=6$ (top row) and $\dot{m}=20$ (bottom row). In this calculation, we fix $M=10^{6}M_{\odot}$, $\alpha=0.3$, $\beta=1$. \textbf{Left panels:} Fraction of energy dissipated in the corona ($f_{\rm c}$) and in the disc ($1-f_{\rm c}$). \textbf{Middle panels:} Radial profiles of temperature ($T/10^9$ K), number density ($n/10^{11}$ cm$^{-3}$), optical depth ($\tau$), Compton $y$-parameter, and Alfv\'en speed ($V_A/c$). \textbf{Right panels:} Emergent spectra corresponding to the two accretion rates. The blue solid curves show the emergent spectra calculated from the new disc-corona model in this paper, the grey dashed curves show the emergent spectra {\color{blue}using} the disc-corona model in \citet{2003ApJ...587..571L}. The light blue shading marks the 0.3-10 keV band.} 
\label{Fig:fig3}
\end{figure*}

Radiation pressure-dominated solution:
Equation~(\ref{Equ:equ14}) can be solved by setting a fixed value
of $c_{1}$. In Fig.~\ref{Fig:fig2}, we plot the mass accretion rate $\dot{m}$ as a function of radius $R/R_{\rm S}$ by setting $c_1(r,\dot{m})=4/27$ for $M=10^{6}M_{\odot}$ and $M=10^{7}M_{\odot}$
respectively. One can see the blue solid line for $M=10^{6}M_{\odot}$ and the orange dashed line for $M=10^{7}M_{\odot}$ respectively in Fig.~\ref{Fig:fig2} for details. 
The two curves correspond to the critical (minimum) mass accretion rate at which a radiation pressure-dominated solution can exist at a fixed radius for 
$M=10^{6}M_{\odot}$ and $M=10^{7}M_{\odot}$ respectively.
We also note that there are minimal values of the curves, which correspond to a mass accretion rate of $\dot m_{\rm min} \sim 0.37$ for $M=10^{6}M_{\odot}$ and $\dot m_{\rm min} \sim 0.29$ for $M=10^{7}M_{\odot}$.
It is clear that if $\dot m> \dot m_{\rm min}$, the radiation pressure-dominated solution can exist. Specifically, for larger $\dot m$,
e.g., $\dot{m}=6$ and $\dot{m}=20$, the size of the radiation dominated solution
is larger than $R_{\rm out}$. While for a medium $\dot m$, e.g., $\dot{m}=1.6$, the accretion will be an inner radiation pressure-dominated solution plus an outer gas pressure-dominated solution.

In Fig.~\ref{Fig:fig3}, we show the fraction of the energy dissipated in the corona $f_{\rm c}$ (left panel), the coronal structures (middle panel) and the emergent spectra (right panel) for $\dot{m}=6$ and $\dot{m}=20$ respectively by fixing $M=10^{6}M_{\odot}$.
It can be seen that in the two cases, the corona is relatively weak, and nearly all the accretion energy is dissipated in the slim disc, i.e., $f_{c}\sim 0$. The spectra are soft in X-rays, dominated by the slim disc. 
For comparison, in the right panel we also plot the emergent spectra calculated using the disc-corona model in \citet{2003ApJ...587..571L} (grey dashed curves). It can be seen that the spectra obtained from our new disc-corona model (blue solid curves) are less luminous in the UV and soft X-ray bands than those of the disc-corona model of \citet{2003ApJ...587..571L} for both $\dot{m}=6$ and $\dot{m}=20$. We also note that the difference of the emergent spectrum between our new disc-corona model and the disc-corona model in \citet{2003ApJ...587..571L} for $\dot{m}=6$ is smaller than that for $\dot{m}=20$. 
This is because the advection effect of the slim disc becomes more significant as the mass accretion rate increases.

\begin{figure*}
\centering
\includegraphics[scale=0.435]{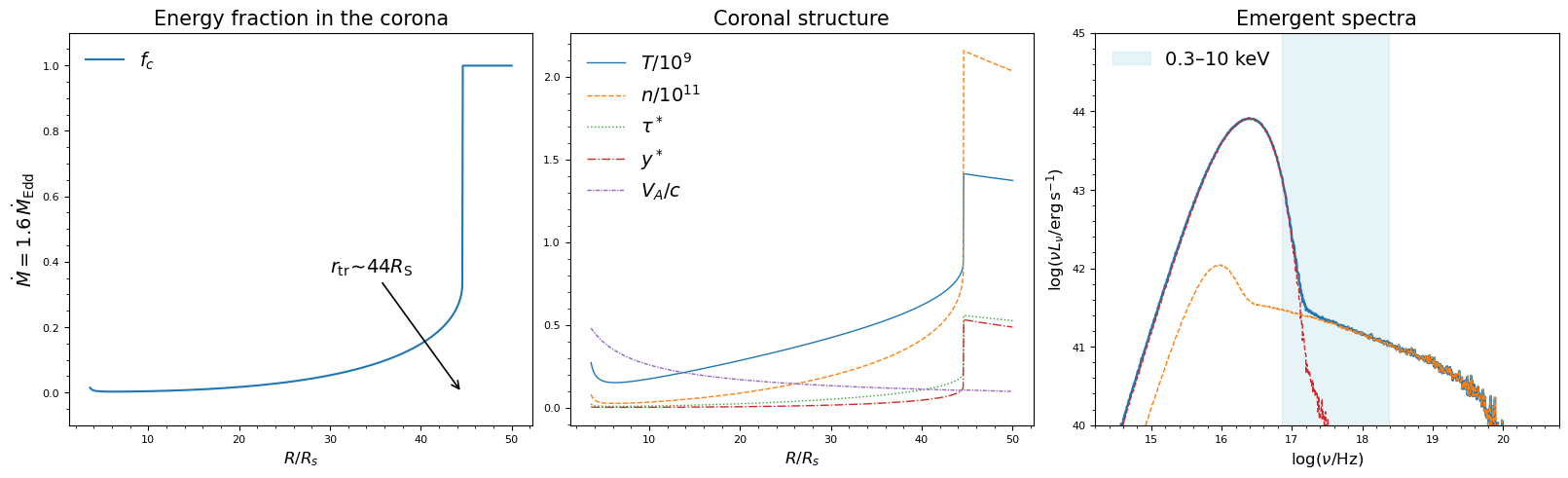}
\caption{Composite solutions (inner radiation pressure-dominated solution + outer gas pressure-dominated solution) for $\dot{m}=1.6$. In this calculation, we fix $M=10^{6}M_{\odot}$, $\alpha=0.3$, $\beta=1$. \textbf{Left panel:} Fraction of energy dissipated in the corona ($f_{\rm c}$). \textbf{Middle panels:} Radial profiles of temperature ($T/10^9$ K), number density ($n/10^{11}$ cm$^{-3}$), optical depth ($\tau$), Compton $y$-parameter, and Alfv\'en speed ($V_A/c$). \textbf{Right panels:} Emergent spectra. The blue solid curve shows the total spectrum. The red dashed curve shows the spectrum from the inner radiation pressure-dominated region, the orange dashed curve shows the spectrum from the outer gas pressure-dominated region. The light blue shading marks the 0.3-10 keV band.}  
\label{Fig:Fig4}
\end{figure*}

Here, we can define the composite solution, which is the inner radiation pressure-dominated solution (dominated by slim disc) + outer gas pressure-dominated solution (disc-corona structure, dominated by corona) for medium $\dot m$.\footnote{For $\dot m$ slightly above $\dot m_{\rm min}$, there is a narrow gas pressure-dominated region between $R_{\rm in}$ and the critical curve of $\dot m$ as a function of  $R/R_{\rm S}$. Since the contribution of this narrow region to the total emission is small, we simply express the structure of accretion flow as an inner radiation pressure-dominated solution + an outer gas pressure-dominated solution, while neglecting the emission from this narrow gas pressure-dominated region in the practical calculations.} We take $M=10^{6}M_{\odot}$ and $\dot{m}=1.6$ as an example, and plot the coronal dissipation fraction $f_{\rm c}$ (left panel), coronal structures (middle panel) and emergent spectra (right panel) in Fig. \ref{Fig:Fig4}. We note that the inner region produces soft spectra dominated by the slim disc radiation and the outer region produces relatively hard spectra dominated by the coronal radiation.

\begin{figure}
\centering
\includegraphics[scale=0.54]{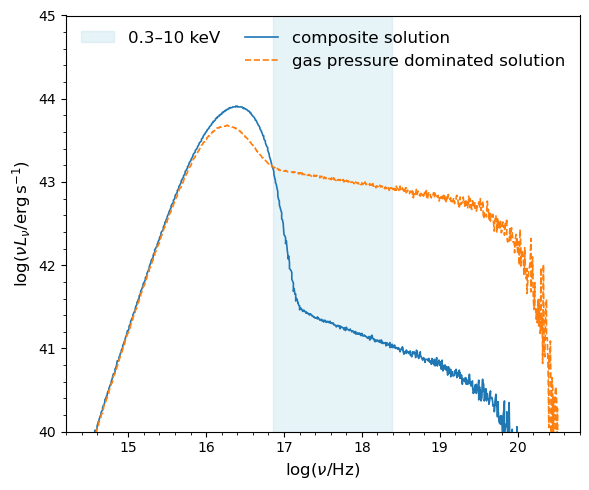}
\caption{Gas pressure-dominated solution (orange dashed line) and composite solutions (blue solid line) for $\dot{m}=1.6$. In this calculation, we fix $M=10^{6}M_{\odot}$, $\alpha=0.3$, $\beta=1$.} 
\label{Fig:fig5}
\end{figure}

In Fig. \ref{Fig:fig5}, we plot the gas pressure-dominated solution and the composite solution for comparison taking $M=10^{6}M_{\odot}$ and $\dot{m}=1.6$. It is clear that the spectrum produced by the composite solution is softer than that of the gas pressure-dominated solution. 

In this paper, we focus on the X-ray spectral evolution of TDEs. Since the early super-Eddington accretion phase, the observed X-ray spectra are very soft and can be well fitted with a blackbody or a multicolour blackbody,
which is inconsistent with the prediction by the gas pressure-dominated solution. 
So we take the composite solutions for matching the observations. 
In this scenario, there is a key quantity, i.e., the transition radius $r_{\rm tr}$, which can control the relative contribution of the emission of the inner radiation pressure-dominated solution and the outer gas pressure-dominated solution. The transition radius $r_{\rm tr}$ can be self-consistently worked out by solving equation~(\ref{Equ:equ14}) for a given $\dot m$ by setting $c_{1}=4/27$, and $r_{\rm tr}$ increases with increasing $\dot m$ as shown in Fig.~\ref{Fig:fig2}. For example, from the left panel of Fig.~\ref{Fig:Fig4} we can clearly see that for $M=10^{6}M_{\odot}$ and $\dot{m}=1.6$, the transition radius $r_{\rm tr}$ is $\sim 44R_{S}$.

\begin{figure*}
\centering
\includegraphics[scale=0.72]{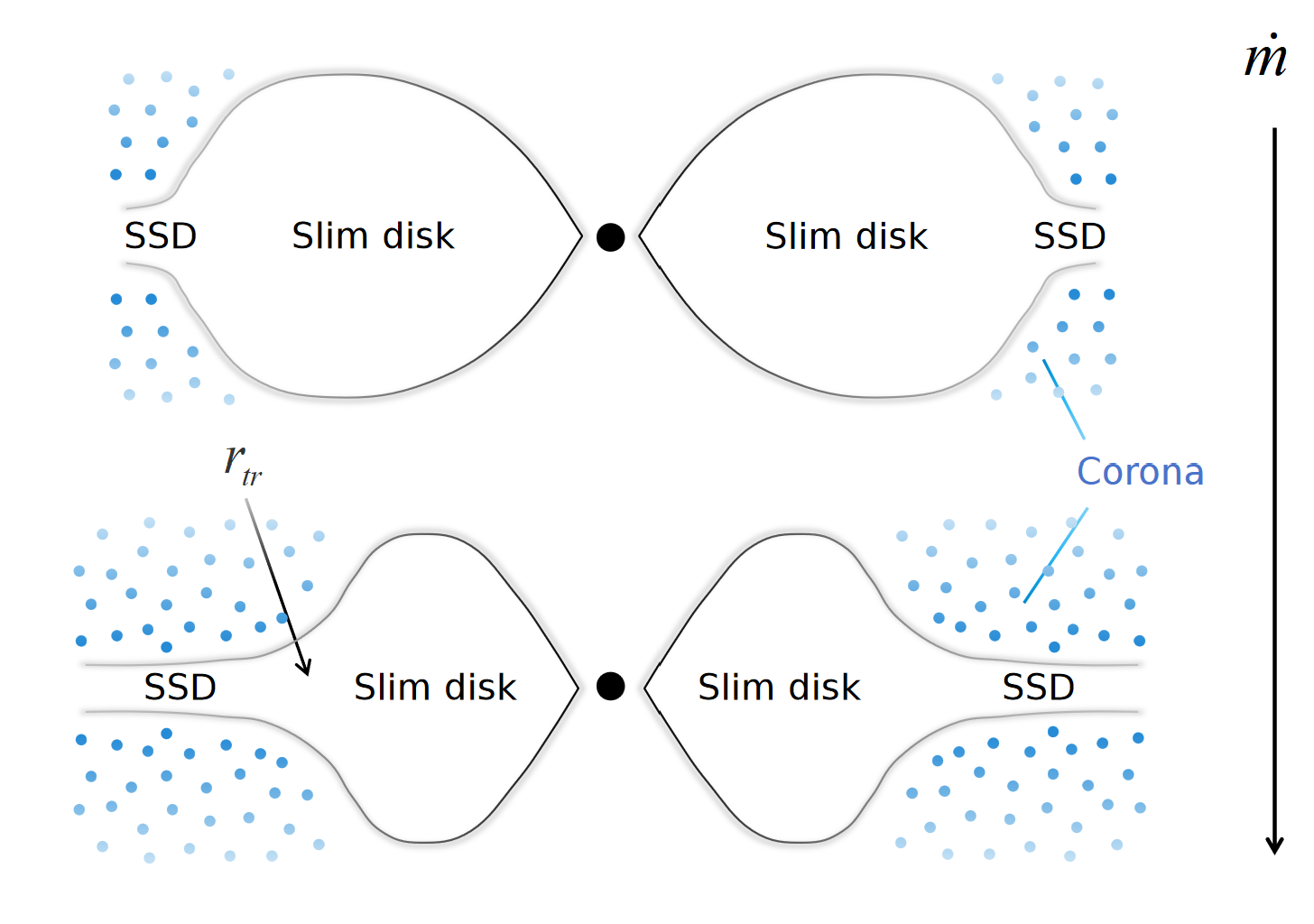}
\caption{The evolution of the geometry of the accretion flow in TDEs with decreasing $\dot m$ from top to bottom.} 
\label{Fig:fig6}

\end{figure*}

Finally, we proposed a scenario for the evolution of the geometry of the accretion flow in TDEs  with decreasing $\dot m$, as shown in Fig.~\ref{Fig:fig6}.

\section{The Application to TDEs} \label{sec:3}

In this paper, we consider a standard case for TDEs.
We assume that a star with mass $M_{\ast}=m_{\ast}M_{\odot}$ and radius $R_{\ast}=r_{\ast}R_{\odot}$ moves in a parabolic orbit around a supermassive BH of mass $M = 10^{6} m_{6} M_{\odot}$.
The star is assumed to be tidally disrupted when it passes close to the pericenter of its orbit, where the tidal force of the BH exceeds the star’s self-gravity \citep{1988Natur.333..523R}. The tidal disruption radius can be expressed as \citep{1975Natur.254..295H, 1988Natur.333..523R},
\begin{equation}
\label{Equ:equ17}
R_{\text{t}} = R_{\ast}\left(\frac{M}{M_{\ast}}\right)^{1/3} \approx 23.7m_{6}^{-2/3}m_{\ast}^{-1/3}r_{\ast}R_{\text{S}}.
\end{equation}

Assuming that the disruption occurs at the pericenter, i.e., $R_{\text{p}}=R_{\text{t}}$, the circularization radius of the debris can be written as,
\begin{equation}
\label{Equ:equ18}
R_{\text{c}} = 2R_{\text{p}} \approx 47m_{6}^{-2/3}m_{\ast}^{-1/3}r_{\ast}R_{\text{S}}.
\end{equation}

The fallback timescale of the TDE is estimated as \citep{2011MNRAS.410..359L},
\begin{equation}
\label{Equ:equ19}
t_{\text{fb}} \approx 41m_{6}^{1/2}m_{\ast}^{-1}r_{\ast}^{3/2}{\rm d}.
\end{equation}

The peak accretion rate can be expressed as,
\begin{equation}
\label{Equ:equ20}
\frac{\dot{M}_{\text{p}}}{\dot{M}_{\text{Edd}}} \approx 133m_{6}^{-3/2}m_{\ast}^{2}r_{\ast}^{-3/2}.
\end{equation}

The classical time-dependent fallback rate of the bound debris follows,
\begin{equation}
\label{Equ:equ21}
\dot{M}_{\text{fb}} = \dot{M}_{\text{p}}\left(\frac{t}{t_{\text{fb}}}+1\right)^{-5/3},\ \ t \geq 0 .
\end{equation}

In this paper, we assume that the mass accretion rate $\dot M$ equals the fallback rate $\dot M_{\rm fb}$. By specifying $M$, $\dot{M}$, $R_{\text{in}}$ and $R_{\text{out}}$ of the accretion disc, as well as $\alpha$ and $\beta$ in the accretion disc, we can use the disc-corona model described in Section \ref {sec:2} to calculate the emergent spectra of TDEs. In the calculations, we fix $\alpha=0.3$ and $\beta=1$ throughout the paper.

\subsection{Emergent Spectra}\label{sec:3.1}

\begin{figure*}
\centering
\includegraphics[scale=0.59]{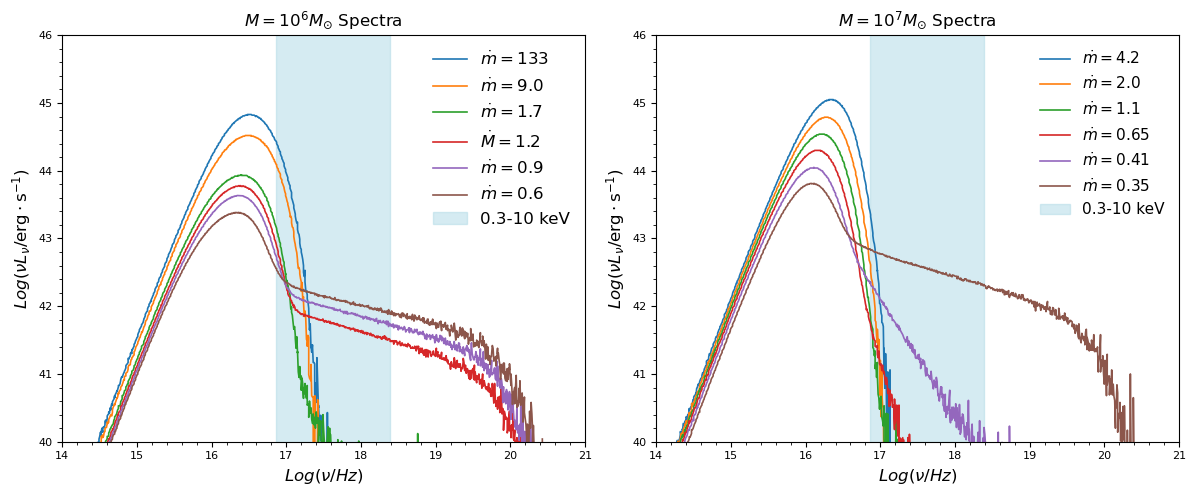}
\caption{\textbf{Left panel}: The variation of the spectrum of a TDE with $M = 10^6 M_{\odot}$ as the accretion rate decreases. The selected accretion rates are $\dot{m} = 133$, $9.0$, $1.7$, $1.2$, $0.9$ and $0.6$. \textbf{Right panel}: The variation of the spectrum of a TDE with $M = 10^7 M_{\odot}$ as the accretion rate decreases. The selected accretion rates are $\dot{m} = 4.2$, $2$, $1.1$, $0.65$, $0.42$ and $0.35$. The 0.3-10 keV X-ray band is highlighted.}
\label{Fig:Fig7}
\end{figure*}

\begin{figure*}
\centering
\includegraphics[scale=0.58]{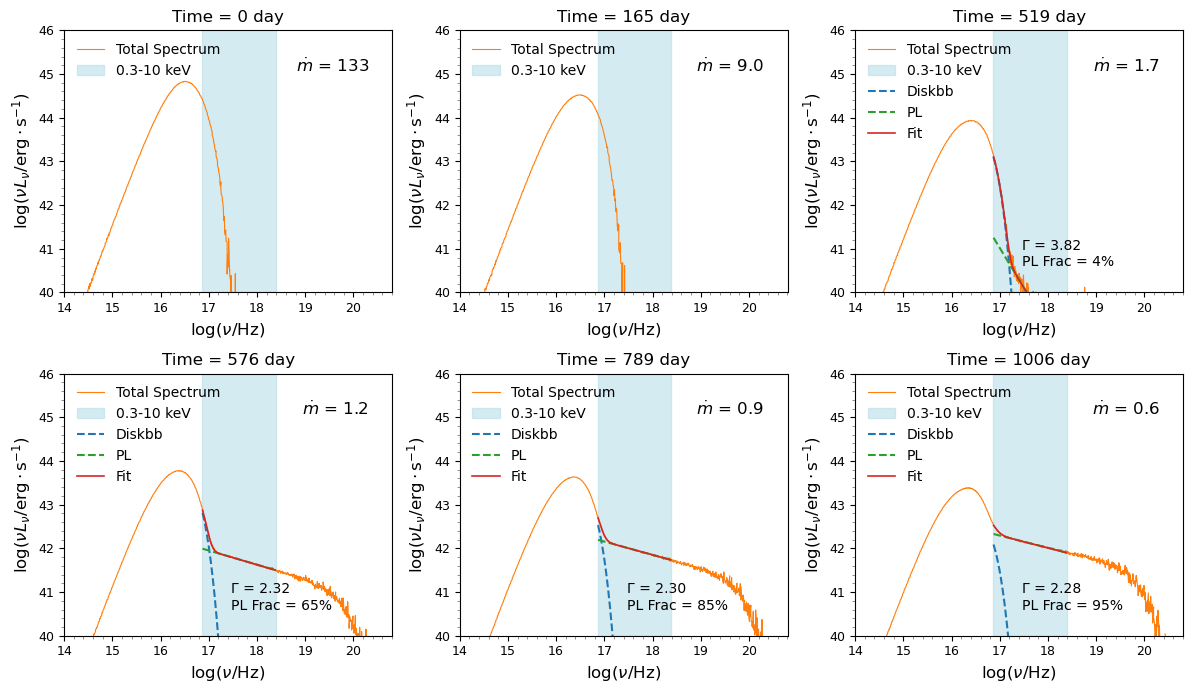}
\caption{Emergent spectra for a TDE with a BH mass of $10^6 M_\odot$ over six epochs. Orange curves are the total emergent spectra of the disc-corona model. The blue and green dashed lines are the best fitting spectra with a \texttt{diskbb} and power-law components for the spectra in 0.3-10 keV. Red curves are \texttt{diskbb} component plus power-law component. Light blue shading marks the 0.3–10 keV band.}
\label{Fig:Fig8}
\end{figure*}

For a BH with a mass of $10^{6}M_{\odot}$, we assume that the disrupted star has solar mass and radius. The inner radius of the accretion disc is set to $R_{\text{in}}=3R_{\text{S}}$, and the outer radius to the circularization radius, from Equation~(\ref{Equ:equ18}) we obtain $R_{\text{out}}=47R_{\text{S}}$. The peak accretion rate for this case, given by Equation~(\ref{Equ:equ20}), is $\dot{M}_{\rm peak}\sim 133\,\dot{M}_{\text{Edd}}$ (i.e. $\dot{m}_{\rm peak}\sim 133$). We then evolve $\dot{M}$ following the fallback law $\dot{M}(t)\propto t^{-5/3}$ (Equation~(\ref{Equ:equ21})) and consider sample values $\dot{m}=133$, $9.0$, $1.7$, $1.2$, $0.9$ and $0.6$, corresponding to $t=0$, $165$, $519$, $576$, $789$ and $1006$ day, respectively. We plot the emergent spectra for different $\dot{m}$ in the left panel of Fig. \ref{Fig:Fig7}.

For a BH with $M = 10^7 M_{\odot}$, the peak accretion rate given by Equation~(\ref{Equ:equ20}) is $\sim 4.2 \dot{M}_{\text{Edd}}$. In this case, the outer radius is set to $R_{\text{out}} = 11R_{\text{S}}$, obtained from Equation~(\ref{Equ:equ18}). Following the $t^{-5/3}$ fallback law, we consider $\dot{m} = 4.2$, $2.0$, $1.1$, $0.65$, $0.42$ and $0.35$, corresponding to $t=0$, $72$, $160$, $267$, $394$ and $446$ day, respectively. We plot the emergent spectra for different $\dot{m}$ in the right panel of Fig. \ref{Fig:Fig7}. We note that as the accretion rate $\dot m$ decreases with time, the X-ray spectrum becomes harder, i.e., evolving from a slim disc dominated state to a disc-corona dominated state. It is clear that the trend of X-ray spectral evolution is consistent with the observations in TDEs.

In order to investigate the details of the spectral evolution with decreasing $\dot m$, we plot the emergent spectra for different $\dot m$ one by one in Fig.~\ref{Fig:Fig8}. 
We fit the X-ray spectra in $0.3\text{--}10~\text{keV}$
with the model of a \texttt{diskbb} (multicolour disc) (\citet{1984PASJ...36..741M,1986ApJ...308..635M,1998PASJ...50..667K})
plus a power-law component. The detailed fitting results can be seen in Table \ref{tab:table1}. 
For $M=10^{6}M_{\odot}$ at $t=0$ day ($\dot m$ = 133) and $t=165$ day ($\dot m$ = 9.0), the X-ray spectra in $0.3\text{--}10~\text{keV}$ are completely
dominated by slim disc, and the power-law (PL) fraction $\sim 0$. In these cases, the temperature at the inner radius $kT_{\rm in}$
are $75.5$ and $67.4\ \rm eV$ respectively. 
With a decrease of $\dot m$ from $1.7$ to $0.6$ corresponding to $t = 519-1006$ days, it can be seen that $kT_{\rm in}$ decreases from $61.1$ to $52.3\ \rm eV$. Meanwhile, the X-ray photon index decreases from $\Gamma=3.82$ to $2.28$, and the PL fraction increases from $3.74\%$ to $94.53\%$.

\begin{figure*}
\centering
\includegraphics[scale=0.58]{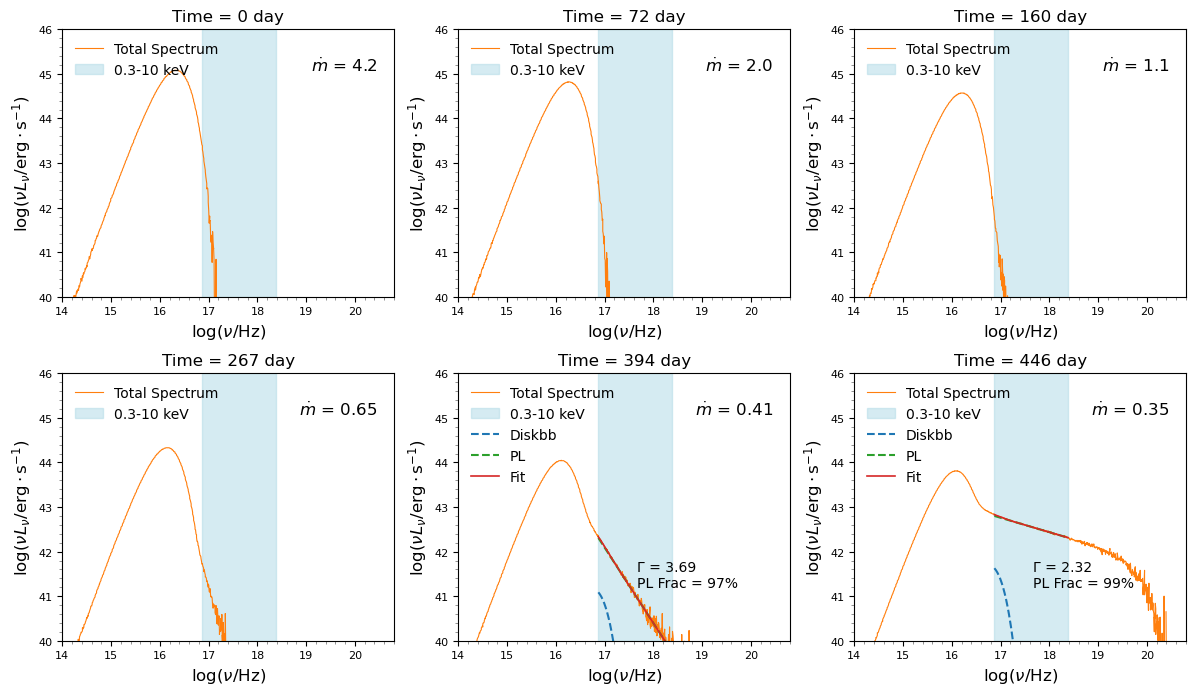}
\caption{Emergent spectra for a TDE with a BH mass of $10^7 M_\odot$ over six epochs. Orange curves are the total emergent spectra of the disc-corona model. The blue and green dashed lines are the best fitting spectra with a \texttt{diskbb} and power-law components for the spectra in 0.3-10 keV. Red curves are \texttt{diskbb} component plus power-law component. Light blue shading marks the 0.3–10 keV band.}
\label{Fig:Fig9}
\end{figure*}

For a BH with mass $M = 10^7M_{\odot}$, the evolutionary time corresponding to $\dot{m} = 4.2$, $2.0$, $1.1$, $0.65$, $0.41$ and $0.35$ are $t=0$, $72$, $160$, $267$, $394$ and $446$ day, respectively. We plot the emergent spectra for different $\dot m$ one by one in Fig.~\ref{Fig:Fig9}. The detailed fitting results can be seen in Table \ref{tab:table1}.
For $M=10^{7}M_{\odot}$, for $t=0$ day ($\dot m$ = 4.2) and $t=72$ day ($\dot m$ = 2.0), the X-ray spectra in $0.3\text{--}10~\text{keV}$ are completely
dominated by slim disc, and the PL fraction $\sim 0$. In these cases, $kT_{\rm in}$
are $36.5$ and $31.4\ \rm eV$ respectively. 
With a decrease of $\dot m$ from $1.1$ to $0.35$ corresponding to $t = 160-446$ days, it can be seen that $kT_{\rm in}$ decreases from $27.7$ to $21.4\ \rm eV$. Meanwhile, the X-ray photon index decreases from $\Gamma=7.22$ to $2.32$, and the PL fraction increases from $10.50\%$ to $98.87\%$.

\begin{figure*}
\centering
\includegraphics[scale=0.58]{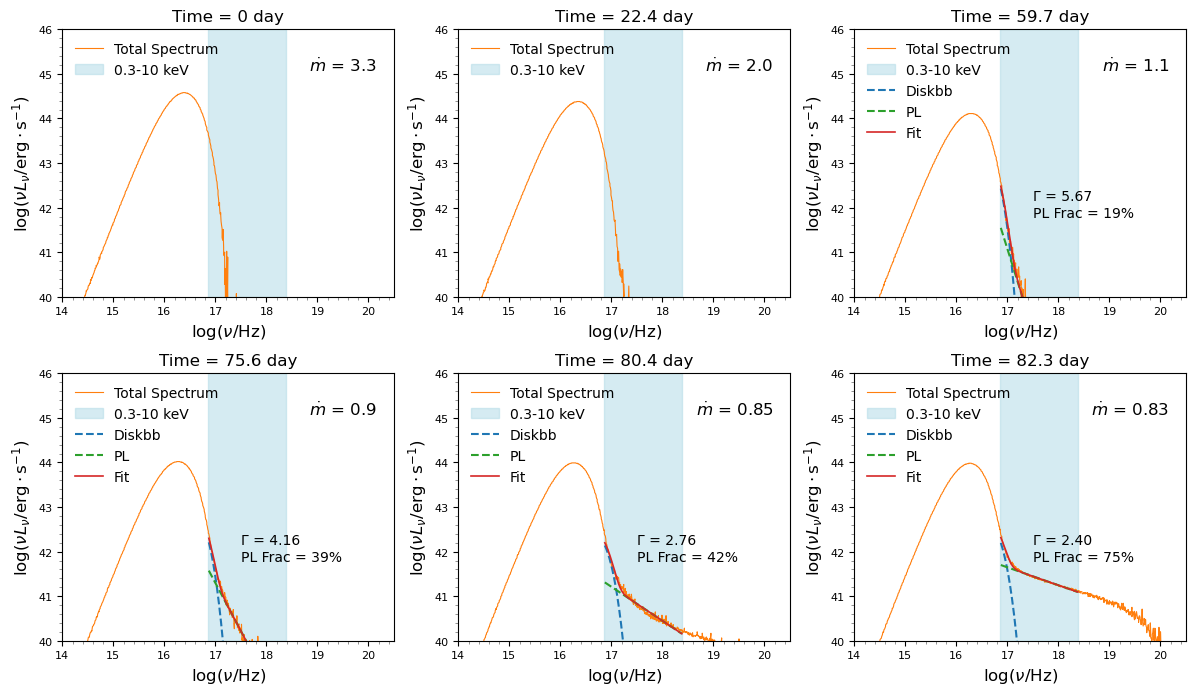}
\caption{Emergent spectra For a TDE with $2.5 \times 10^{6} M_\odot$ (consistent with AT 2019azh) over six epochs. Orange curves are the total emergent spectra of the disc-corona model. The blue and green dashed lines are the best fitting spectra with a \texttt{diskbb} and power-law components for the spectra in 0.3-10 keV. Red curves are \texttt{diskbb} component plus power-law component. Light blue shading marks the 0.3–10 keV band.}
\label{Fig:Fig10}
\end{figure*}

\begin{table}
\caption{Results of spectral fitting}
\label{tab:table1}
\hspace{-0.034\textwidth} 
\resizebox{0.52\textwidth}{!}{
\begin{tabular}{cccccc}
\toprule
\midrule               
$M$ & t (day) & $\dot{m}$ & $kT_{\text{in}}$ (eV) & $\Gamma$ & PL frac (\%) \\
\midrule                      
\multirow{6}{*}{\parbox[c]{1.6cm}{\centering $10^{6}M_{\odot}$}} 
& 0    & 133   & 75.5 & --   & $\sim$0 \\
& 165  & 9.0   & 67.4 & --   & $\sim$0 \\
& 519  & 1.7   & 61.1 & 3.82 & 3.74   \\
& 576  & 1.2   & 58.2 & 2.32 & 65.29  \\
& 780  & 0.9   & 56.3 & 2.30 & 85.22  \\
& 1006 & 0.6   & 52.3 & 2.28 & 94.53  \\

\midrule 

\multirow{6}{*}{\parbox[c]{1.6cm}{\centering $10^{7}M_{\odot}$}}
& 0   & 4.2  & 36.5 & --   & $\sim$0 \\
& 72  & 2.0  & 31.4 & --   & $\sim$0 \\
& 160 & 1.1  & 27.7 & 7.22 & 10.50  \\
& 267 & 0.65 & 24.1 & 5.68 & 62.52  \\
& 394 & 0.41 & 23.1 & 3.69 & 96.55  \\
& 446 & 0.35 & 21.4 & 2.32 & 98.87  \\

\bottomrule
\end{tabular}
}
\\[3pt]
{\bf Note.} $T_{\rm in}$ is the temperature at inner disc radius. $\Gamma$ is the photon index of the power-law component. 
PL frac (\%) is the fraction of the power-law component relative to the total flux in the 0.3--10 keV band. We set $R_{\rm in}=3R_{\rm S}$, $R_{\rm out}=47R_{\rm S}$ for $M=10^{6}M_{\odot}$ and $R_{\rm in}=3R_{\rm S}$, $R_{\rm out}=11R_{\rm S}$ for $M=10^{7}M_{\odot}$.
\end{table}

\subsection{The Application to AT 2019azh}\label{sec:3.2}
\begin{figure*}
\centering
\includegraphics[scale=0.74]{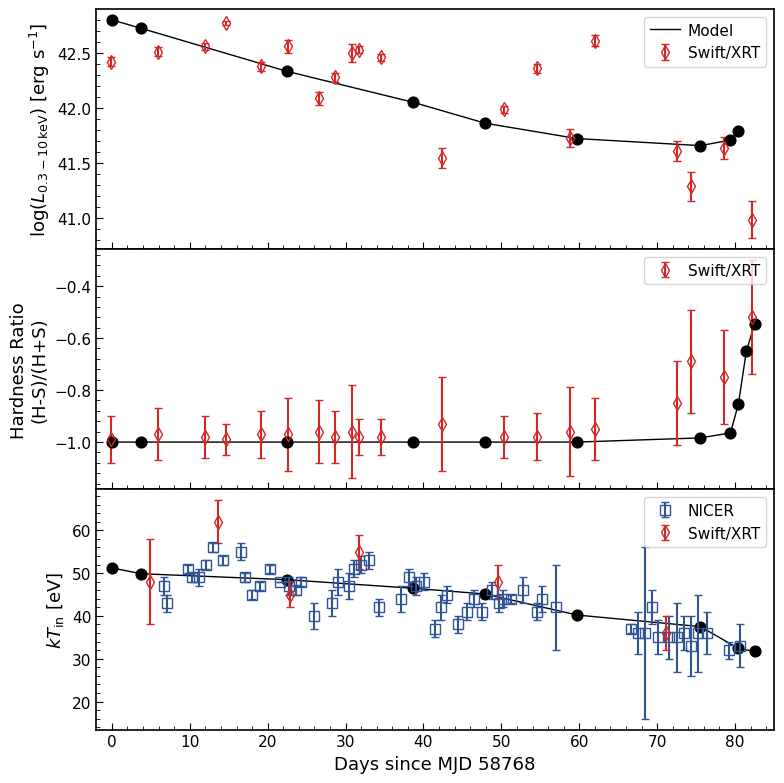}
\caption{Evolution of AT~2019azh. 
\textbf{Top row:} Evolution of 0.3--10\,keV X-ray luminosity (red diamonds: \textit{Swift}/XRT data, black solid line with circles: model). 
\textbf{Middle row:} Evolution of Hardness ratio HR $=$ (H--S)/(H+S), with H and S the number of counts in the 2--10\,keV and 0.3--2\,keV bands (same markers as top row). 
\textbf{Bottom row:} Evolution of Thermal temperature (blue squares: NICER data, red diamonds: \textit{Swift}/XRT data, black solid line with filled circles: model). Error bars show uncertainties. The horizontal axis gives days since MJD~58768.}
\label{Fig:Fig11}

\end{figure*}

AT 2019azh is a TDE candidate discovered on 2019 February 22 by the All-Sky Automated Survey for Supernovae (ASAS-SN; \citet{2014ApJ...788...48S}).
The BH mass of AT 2019azh is estimated to be $2.5 \times 10^{6}M_{\odot}$ \citep{2024ApJ...969..104F}. We match the X-ray observations of 
AT 2019azh (from \citet{2021MNRAS.500.1673H}) by indicating $\dot M(t)$, $R_{\rm in}$ and $R_{\rm out}$ of the accretion disc. Specifically, we take $\dot M=\dot M_{\rm fb}=\dot{M}_{\text{ini}}\left(1+\frac{t}{t_{\text{fb}}}\right)^{-5/3},(t \geq 0)$ for the evolution of the mass accretion rate, where $t_{\rm fb}$ is calculated from equation~(\ref{Equ:equ19}) assuming a solar-type star with a parabolic orbit being disrupted, and $\dot M_{\rm ini}$
is a free parameter compared with equation~(\ref{Equ:equ20}) since it is possible that a fraction of the fallback material could be blown off by shocks generated in the stream-stream collision during circularization \citep{2016ApJ...830..125J}. So in general, $\dot M_{\text{ini}}$ is less than $\dot M_{\text{peak}}$. In this paper,
we take $\dot{M}_{\rm ini}=3.3\dot{M}_{\rm Edd}$ (a discussion of the effect of $\dot{M}_{\rm ini}$ will be shown later.)
Here $R_{\rm in}$ is fixed to be $3R_{\rm S}$ for a non-rotating BH. $R_{\rm out}$ is calculated to be $26R_{\rm S}$ according to equation~(\ref{Equ:equ18}). 
In Fig.~\ref{Fig:Fig10}, we plot the emergent spectra for $t=0$, $22.4$, $59.7$, $75.6$, $80.4$, $82.3$
day corresponding to $\dot m = 3.3$, $2.0$, $1.1$, $0.9$, $0.85$, $0.83$ respectively. 
As in Section \ref{sec:3.1}, we fit the spectra with the model of a \texttt{diskbb} (multicolour disc) component plus a power-law component, and summarize the specific spectral parameters in Table \ref{tab:table2}. To more clearly compare the theoretical results and the observations, we plot the 0.3-10 keV X-ray luminosity, hardness ratio (HR) \footnote{HR $=$ (H--S)/(H+S), where S and H are the count rates in the 0.3–2 keV and 2–10 keV X-ray bands, respectively. We converted the theoretical luminosities in both bands into observed fluxes. These fluxes were then transformed into photon counts using $\text{WebPIMMS}$ (\url{https://heasarc.gsfc.nasa.gov/cgi-bin/Tools/w3pimms/w3pimms.pl}), yielding S and H.} and temperature at inner disc radius as a function of time \footnote{Our theoretical model intrinsically produces a multicolour blackbody spectrum, so fitting with \texttt{diskbb} is more accurate. The observations are fitted with a single blackbody, this difference may introduce some small errors, but the impact is minor due to the relatively small disc size.} in Fig.~\ref{Fig:Fig11}. It can be seen that the theoretical results from our model can match the observations very well. 

\begin{table}
\caption{The theoretical spectral parameters for AT2019azh}
\label{tab:table2}
\hspace{-0.03\textwidth} 
\resizebox{0.51\textwidth}{!}{ 
\begin{tabular}{ccccccc}
\toprule
\midrule                      
t & $\dot{m}$ & $kT_{\text{in}}$ & $\Gamma$ & PL frac & HR & $\log L_{\rm X}$ \\
(day) &  & (eV) &  & (\%) &  & (erg s$^{-1}$) \\
\midrule                      

0    & 3.3   & 51.2 & --   & $\sim$0 & -1 & 42.80\\
22.4  & 2.0   & 48.4 & --   & $\sim$0 & -1 & 42.33\\
59.7  & 1.1   & 40.2 & 5.67 & 19.39  & -0.999 & 41.72\\
75.6  & 0.9   & 37.5 & 4.16 & 39.12  & -0.983 & 41.66\\
80.4  & 0.85  & 32.4 & 2.76 & 41.76  & -0.854 & 41.79\\
82.3 & 0.83   & 31.9 & 2.40 & 74.53  & -0.547 & 41.97\\

\bottomrule
\end{tabular}
}
\\[3pt]
{\bf Note.} $T_{\rm in}$ is the temperature at inner disc radius. $\Gamma$ is the photon index of the power-law component. PL frac (\%) is the fraction of the power-law component relative to the total flux in the 0.3--10 keV band. The hardness ratio (HR) is defined as (H–S)/(H+S), where H and S are the count rates in the 2–10 keV and 0.3–2 keV ranges, respectively. $L_{\rm X}$ is the X-ray luminosity in the 0.3--10 keV band. We fix $M=2.5 \times 10^{6}M_{\odot}$, $R_{\rm in}=3R_{\rm S}$ and $R_{\rm out}=26R_{\rm S}$.
\end{table}

Here, we should note that $\dot{M}_{\rm ini}$ is actually a free parameter
in our model, we test the effect of different $\dot{M}_{\rm ini}$ to the spectral evolution. Following \citet{2021MNRAS.500.1673H}, we set $t=0$ at the first Swift/XRT detection of significant X-ray emission ($\sim$ 225 days after the optical/UV peak). We note that adopting a different $\dot{M}_{\rm ini}$ would affect the evolution timescale: a higher $\dot{M}_{\rm ini}$ generally leads to a slower evolution, while a lower $\dot{M}_{\rm ini}$ would result in a faster decline. By testing several values, we find that $\dot{M}_{\rm ini}=3.3\dot{M}_{\rm Edd}$ matches the observations well.

\section{Summary and discussion} \label{sec:4}
In this paper, we construct a new disc-corona model to explain the X-ray spectral evolution in TDEs. Specifically, we replaced the standard accretion disc by a slim disc for the structure
of the disc and the corona, based on which we further calculated the emergent spectra of the accretion flow with the method of Monte Carlo simulations. We then proposed a scenario of the geometry of the accretion flow in TDEs, i.e., an inner slim disc plus an outer disc-corona system as in Fig. \ref{Fig:fig6}. There is a transition radius $r_{\rm tr}$ between these two regions. Our study shows that $r_{\rm tr}$ decreases with decreasing $\dot{M}$, which can predict the hardening of the X-ray spectra since the relative contribution of the hard components increases while that of the soft components decreases with decreasing $\dot {M}$. Our model has been successfully applied to explain  the spectral evolution of the TDE candidate AT 2019azh, including the X-ray luminosity, the hardness ratio, and the temperature at inner disc radius as a function of time. We expect that our model can be 
used to explain the X-ray observations for more TDEs in the future. This includes two main aspects,
i.e., the improvement to the model and the applications to new observational data.


As for the new disc-corona model in the present paper, we still separately treat the radiation pressure-dominated and the gas pressure-dominated accretion disc for the structure of the disc and the corona as that in \citet{2003ApJ...587..571L}, which will lead to an abrupt change of the structure of the corona, such as electron temperature, number density, effective optical depth, Compton $y$ parameter and Alfv\'en speed in Fig. \ref{Fig:Fig4}. In the future, we expect to combine
the effects of radiation pressure and gas pressure together to recalculate the structure of the disc and the corona, so that we can theoretically obtain a solution of the disc and the corona smoothly. Meanwhile, we should note that in the present paper, we only very simply match the X-ray 
observational data with our disc-corona model, which actually is not a strict fitting. 
So in this framework of the smooth solution we mentioned above, we try to test model parameters such as BH mass $M$, accretion rate $\dot{M}$, viscosity $\alpha$, magnetic $\beta$ to the disc-corona structure and the emergent spectra. Further, we try to construct a database of the emergent spectra for different model parameters to precisely fit the X-ray observations. 

In addition, as we can see in the present paper, we use the stationary disc-corona model to explain the X-ray spectral
evolution in TDEs. Specifically, the change of the disc-corona structure is resulted by changing $\dot{M}$. 
We argue this approximation is reasonable. This is because the equilibrium timescale between the disc and the corona is the thermal timescale, which generally can be expressed as
$t_{\rm th}\sim \alpha^{-1} \Omega^{-1}$. The viscous timescale for
disc is $t_{\rm vis}\sim \alpha^{-1} \Omega^{-1} (H/R)^{-2}$. When the accretion is at a slim disc dominated phase, the strength of the corona is very weak. As the accretion drops to a sub-Eddington phase, the strength of the corona becomes important. In this case, the viscous timescale $t_{\rm vis}$ is much longer than the thermal timescale $t_{\rm th}$ due to 
$H/R<<1$, so there will be enough time for the equilibrium between the disc and the corona to be re-established if $\dot{M}$ is changed. 
Moreover, in TDEs the evolution of $\dot{M}$ is governed by the fallback rate of the stellar debris. The characteristic timescale for the change of $\dot{M}$ is the fallback timescale $t_{\rm fb}$ (see equation~(\ref{Equ:equ19})). For typical parameters, this corresponds to a timescales of several tens of days. By contrast, the thermal timescale $t_{\rm th}$ is from hours to days in order of magnitude, which is much shorter than $t_{\rm fb}$. 
Therefore, the disc and corona can adjust much faster than the change of $\dot{M}$, supporting that the use of a stationary approximation in the present paper is reasonable.
We should note that, although the stationary disc-corona model presented in the present paper can well match the observations to some extent, it is still very necessary to develop a time-dependent disc-corona for more precisely studying the accretion physics in TDEs. 
Currently, we have developed one dimension 
time-dependent disc model to explain the evolution of
optical/UV emission in TDEs \citep{2025arXiv250916544G}. 
The consideration of coronal component in the time-dependent disc will be carried out in the near future.

We also note that in the present paper, we only consider the emission of the disc and the corona itself, which can explain the X-ray emission in TDEs, however, intrinsically can not explain the optical/UV emission. In general, the optical/UV emission can be either from the reprocessing model where the X-ray emission is reprocessed into the optical band by a surrounding optically thick envelope or outflow, or from the collisions of the debris of the star after the disruption. 
In the framework of the accretion scenario, we expect that the effect of outflow can be added to our
disc-corona model. Generally speaking, it is difficult to consider the outflow with the method of
semi-analytical numerical calculations. We are trying to add the outflow by hand, based on our results of the properties of the outflow from radiation hydrodynamic simulation in TDEs as in \citet{2025MNRAS.539.3473Q}.

Finally, combining the improved disc-corona model with outflows and the upcoming rich multi-band
observational data for TDEs, we expect that we can deeply explore the accretion physics in TDEs.



\section*{Acknowledgements}

Wei Chen thanks the useful discussions with Chenlei Guo, Jieying Liu, Mingjun Liu, Yiyang Lin and Yongxin Wu. We also thank the
organizers and participants of the TDE Full Process Simulation Seminar Series for the valuable lectures and discussions. This work is supported by the National Key R\&D Program of China (no.2023YFA1607903), the Strategic Priority Research Programme of the Chinese Academy of Science (grant no. XDB0550200) and National Natural Science Foundation of
China (grant no. 12173048, 12333004).

\section*{Data Availability}

The data underlying this article will be shared on reasonable request 
to the corresponding author.



\bibliographystyle{mnras}
\bibliography{references} 

@ARTICLE{1975Natur.254..295H,
       author = {{Hills}, J.~G.},
        title = "{Possible power source of Seyfert galaxies and QSOs}",
      journal = {\nat},
         year = 1975,
        month = mar,
       volume = {254},
       number = {5498},
        pages = {295-298},
          doi = {10.1038/254295a0},
       adsurl = {https://ui.adsabs.harvard.edu/abs/1975Natur.254..295H}
}

@ARTICLE{1988Natur.333..523R,
       author = {{Rees}, Martin J.},
        title = "{Tidal disruption of stars by black holes of {}10$^{6}$-{}10$^{8}$ solar masses in nearby galaxies}",
      journal = {\nat},
         year = 1988,
        month = jun,
       volume = {333},
       number = {6173},
        pages = {523-528},
          doi = {10.1038/333523a0},
       adsurl = {https://ui.adsabs.harvard.edu/abs/1988Natur.333..523R}
}

@ARTICLE{1988ApJ...332..646A,
       author = {{Abramowicz}, M.~A. and {Czerny}, B. and {Lasota}, J.~P. and {Szuszkiewicz}, E.},
        title = "{Slim Accretion Disks}",
      journal = {\apj},
         year = 1988,
        month = sep,
       volume = {332},
        pages = {646},
          doi = {10.1086/166683},
       adsurl = {https://ui.adsabs.harvard.edu/abs/1988ApJ...332..646A}
}

@ARTICLE{2016MNRAS.458.2454L,
       author = {{Lubi{\'n}ski}, P. and {Beckmann}, V. and {Gibaud}, L. and {Paltani}, S. and {Papadakis}, I.~E. and {Ricci}, C. and {Soldi}, S. and {T{\"u}rler}, M. and {Walter}, R. and {Zdziarski}, A.~A.},
        title = "{A comprehensive analysis of the hard X-ray spectra of bright Seyfert galaxies}",
      journal = {\mnras},
         year = 2016,
        month = may,
       volume = {458},
       number = {3},
        pages = {2454-2475},
          doi = {10.1093/mnras/stw454},
archivePrefix = {arXiv},
       eprint = {1602.08402},
 primaryClass = {astro-ph.GA},
       adsurl = {https://ui.adsabs.harvard.edu/abs/2016MNRAS.458.2454L}
}

@ARTICLE{2000ApJ...542..703Z,
       author = {{Zdziarski}, Andrzej A. and {Poutanen}, Juri and {Johnson}, W. Neil},
        title = "{Observations of Seyfert Galaxies by OSSE and Parameters of Their X-Ray/Gamma-Ray Sources}",
      journal = {\apj},
         year = 2000,
        month = oct,
       volume = {542},
       number = {2},
        pages = {703-709},
          doi = {10.1086/317046},
archivePrefix = {arXiv},
       eprint = {astro-ph/0006151},
 primaryClass = {astro-ph},
       adsurl = {https://ui.adsabs.harvard.edu/abs/2000ApJ...542..703Z}
}

@ARTICLE{1998MNRAS.301..179M,
       author = {{Magdziarz}, Pawel and {Blaes}, Omer M. and {Zdziarski}, Andrzej A. and {Johnson}, W. Neil and {Smith}, David A.},
        title = "{A spectral decomposition of the variable optical, ultraviolet and X-ray continuum of NGC 5548}",
      journal = {\mnras},
         year = 1998,
        month = nov,
       volume = {301},
       number = {1},
        pages = {179-192},
          doi = {10.1046/j.1365-8711.1998.02015.x},
       adsurl = {https://ui.adsabs.harvard.edu/abs/1998MNRAS.301..179M}
}

@ARTICLE{1999ApJ...514..180U,
       author = {{Ulmer}, Andrew},
        title = "{Flares from the Tidal Disruption of Stars by Massive Black Holes}",
      journal = {\apj},
         year = 1999,
        month = mar,
       volume = {514},
       number = {1},
        pages = {180-187},
          doi = {10.1086/306909},
       adsurl = {https://ui.adsabs.harvard.edu/abs/1999ApJ...514..180U}
}

@INPROCEEDINGS{1989IAUS..136..543P,
       author = {{Phinney}, E.~S.},
        title = "{Manifestations of a Massive Black Hole in the Galactic Center}",
    booktitle = {The Center of the Galaxy},
         year = 1989,
       editor = {{Morris}, Mark},
       series = {IAU Symposium},
       volume = {136},
        month = jan,
        pages = {543},
       adsurl = {https://ui.adsabs.harvard.edu/abs/1989IAUS..136..543P}
}

@ARTICLE{2011MNRAS.410..359L,
       author = {{Lodato}, Giuseppe and {Rossi}, Elena M.},
        title = "{Multiband light curves of tidal disruption events}",
      journal = {\mnras},
         year = 2011,
        month = jan,
       volume = {410},
       number = {1},
        pages = {359-367},
          doi = {10.1111/j.1365-2966.2010.17448.x},
archivePrefix = {arXiv},
       eprint = {1008.4589},
 primaryClass = {astro-ph.CO},
       adsurl = {https://ui.adsabs.harvard.edu/abs/2011MNRAS.410..359L}
}

@ARTICLE{1973A&A....24..337S,
       author = {{Shakura}, N.~I. and {Sunyaev}, R.~A.},
        title = "{Black holes in binary systems. Observational appearance.}",
      journal = {\aap},
         year = 1973,
        month = jan,
       volume = {24},
        pages = {337-355},
       adsurl = {https://ui.adsabs.harvard.edu/abs/1973A&A....24..337S}
}

@MISC{2021SSRv..217...18S,
       author = {{Saxton}, R. and {Komossa}, S. and {Auchettl}, K. and {Jonker}, P.~G.},
        title = "{Correction to: X-Ray Properties of TDEs}",
 howpublished = {Space Science Reviews, Volume 217, Issue 1, article id.18},
         year = 2021,
        month = feb,
          eid = {18},
        pages = {18},
          doi = {10.1007/s11214-020-00759-7},
archivePrefix = {arXiv},
       eprint = {2103.15442},
 primaryClass = {astro-ph.HE},
       adsurl = {https://ui.adsabs.harvard.edu/abs/2021SSRv..217...18S}
}

@ARTICLE{2004ApJ...603L..17K,
       author = {{Komossa}, Stefanie and {Halpern}, Jules and {Schartel}, Norbert and {Hasinger}, G{\"u}nther and {Santos-Lleo}, Maria and {Predehl}, Peter},
        title = "{A Huge Drop in the X-Ray Luminosity of the Nonactive Galaxy RX J1242.6-1119A, and the First Postflare Spectrum: Testing the Tidal Disruption Scenario}",
      journal = {\apjl},
         year = 2004,
        month = mar,
       volume = {603},
       number = {1},
        pages = {L17-L20},
          doi = {10.1086/382046},
archivePrefix = {arXiv},
       eprint = {astro-ph/0402468},
 primaryClass = {astro-ph},
       adsurl = {https://ui.adsabs.harvard.edu/abs/2004ApJ...603L..17K}
}

@ARTICLE{1999A&A...343..775K,
       author = {{Komossa}, Stefanie and {Bade}, Norbert},
        title = "{The giant X-ray outbursts in NGC 5905 and IC 3599:() hfill Follow-up observations and outburst scenarios}",
      journal = {\aap},
         year = 1999,
        month = mar,
       volume = {343},
        pages = {775-787},
          doi = {10.48550/arXiv.astro-ph/9901141},
archivePrefix = {arXiv},
       eprint = {astro-ph/9901141},
 primaryClass = {astro-ph},
       adsurl = {https://ui.adsabs.harvard.edu/abs/1999A&A...343..775K}
}

@ARTICLE{2014ApJ...792L..29M,
       author = {{Maksym}, W. Peter and {Lin}, Dacheng and {Irwin}, Jimmy A.},
        title = "{RBS 1032: A Tidal Disruption Event in Another Dwarf Galaxy?}",
      journal = {\apjl},
         year = 2014,
        month = sep,
       volume = {792},
       number = {2},
          eid = {L29},
        pages = {L29},
          doi = {10.1088/2041-8205/792/2/L29},
archivePrefix = {arXiv},
       eprint = {1407.2928},
 primaryClass = {astro-ph.HE},
       adsurl = {https://ui.adsabs.harvard.edu/abs/2014ApJ...792L..29M}
}

@ARTICLE{2021ApJ...912..151W,
       author = {{Wevers}, T. and {Pasham}, D.~R. and {van Velzen}, S. and {Miller-Jones}, J.~C.~A. and {Uttley}, P. and {Gendreau}, K.~C. and {Remillard}, R. and {Arzoumanian}, Z. and {L{\"o}wenstein}, M. and {Chiti}, A.},
        title = "{Rapid Accretion State Transitions following the Tidal Disruption Event AT2018fyk}",
      journal = {\apj},
         year = 2021,
        month = may,
       volume = {912},
       number = {2},
          eid = {151},
        pages = {151},
          doi = {10.3847/1538-4357/abf5e2},
archivePrefix = {arXiv},
       eprint = {2101.04692},
 primaryClass = {astro-ph.HE},
       adsurl = {https://ui.adsabs.harvard.edu/abs/2021ApJ...912..151W}
}

@ARTICLE{2024ApJ...970..116W,
       author = {{Wen}, S. and {Jonker}, P.~G. and {Levan}, A.~J. and {Li}, D. and {Stone}, N.~C. and {Zabludoff}, A.~I. and {Cao}, Z. and {Wevers}, T. and {Pasham}, D.~R. and {Lewin}, C. and {Kara}, E.},
        title = "{AT2018fyk: Candidate Tidal Disruption Event by a (Super)Massive Black Hole Binary}",
      journal = {\apj},
         year = 2024,
        month = aug,
       volume = {970},
       number = {2},
          eid = {116},
        pages = {116},
          doi = {10.3847/1538-4357/ad4da3},
archivePrefix = {arXiv},
       eprint = {2405.00894},
 primaryClass = {astro-ph.HE},
       adsurl = {https://ui.adsabs.harvard.edu/abs/2024ApJ...970..116W}
}

@ARTICLE{2023MNRAS.519.2375C,
       author = {{Cao}, Z. and {Jonker}, P.~G. and {Wen}, S. and {Stone}, N.~C. and {Zabludoff}, A.~I.},
        title = "{The rapidly spinning intermediate-mass black hole 3XMM J150052.0+015452}",
      journal = {\mnras},
         year = 2023,
        month = feb,
       volume = {519},
       number = {2},
        pages = {2375-2390},
          doi = {10.1093/mnras/stac3539},
archivePrefix = {arXiv},
       eprint = {2211.16936},
 primaryClass = {astro-ph.HE},
       adsurl = {https://ui.adsabs.harvard.edu/abs/2023MNRAS.519.2375C}
}

@ARTICLE{2021MNRAS.504.4730M,
       author = {{Mummery}, Andrew and {Balbus}, Steven A.},
        title = "{Hard X-ray emission from a Compton scattering corona in large black hole mass tidal disruption events}",
      journal = {\mnras},
         year = 2021,
        month = jul,
       volume = {504},
       number = {4},
        pages = {4730-4742},
          doi = {10.1093/mnras/stab1184},
archivePrefix = {arXiv},
       eprint = {2104.06195},
 primaryClass = {astro-ph.HE},
       adsurl = {https://ui.adsabs.harvard.edu/abs/2021MNRAS.504.4730M}
}

@ARTICLE{1993ApJ...413..507H,
       author = {{Haardt}, Francesco and {Maraschi}, Laura},
        title = "{X-Ray Spectra from Two-Phase Accretion Disks}",
      journal = {\apj},
         year = 1993,
        month = aug,
       volume = {413},
        pages = {507},
          doi = {10.1086/173020},
       adsurl = {https://ui.adsabs.harvard.edu/abs/1993ApJ...413..507H}
}

@ARTICLE{1991ApJ...380L..51H,
       author = {{Haardt}, F. and {Maraschi}, L.},
        title = "{A Two-Phase Model for the X-Ray Emission from Seyfert Galaxies}",
      journal = {\apjl},
         year = 1991,
        month = oct,
       volume = {380},
        pages = {L51},
          doi = {10.1086/186171},
       adsurl = {https://ui.adsabs.harvard.edu/abs/1991ApJ...380L..51H}
}

@ARTICLE{1996ApJ...470..249P,
       author = {{Poutanen}, Juri and {Svensson}, Roland},
        title = "{The Two-Phase Pair Corona Model for Active Galactic Nuclei and X-Ray Binaries: How to Obtain Exact Solutions}",
      journal = {\apj},
         year = 1996,
        month = oct,
       volume = {470},
        pages = {249},
          doi = {10.1086/177865},
archivePrefix = {arXiv},
       eprint = {astro-ph/9605073},
 primaryClass = {astro-ph},
       adsurl = {https://ui.adsabs.harvard.edu/abs/1996ApJ...470..249P}
}

@ARTICLE{1994ApJ...436..599S,
       author = {{Svensson}, Roland and {Zdziarski}, Andrzej A.},
        title = "{Black Hole Accretion Disks with Coronae}",
      journal = {\apj},
         year = 1994,
        month = dec,
       volume = {436},
        pages = {599},
          doi = {10.1086/174934},
       adsurl = {https://ui.adsabs.harvard.edu/abs/1994ApJ...436..599S}
}

@ARTICLE{2003ApJ...587..571L,
       author = {{Liu}, B.~F. and {Mineshige}, S. and {Ohsuga}, K.},
        title = "{Spectra from a Magnetic Reconnection-heated Corona in Active Galactic Nuclei}",
      journal = {\apj},
         year = 2003,
        month = apr,
       volume = {587},
       number = {2},
        pages = {571-579},
          doi = {10.1086/368282},
archivePrefix = {arXiv},
       eprint = {astro-ph/0301142},
 primaryClass = {astro-ph},
       adsurl = {https://ui.adsabs.harvard.edu/abs/2003ApJ...587..571L}
}

@ARTICLE{2002ApJ...572L.173L,
       author = {{Liu}, B.~F. and {Mineshige}, S. and {Shibata}, K.},
        title = "{A Simple Model for a Magnetic Reconnection-heated Corona}",
      journal = {\apjl},
         year = 2002,
        month = jun,
       volume = {572},
       number = {2},
        pages = {L173-L176},
          doi = {10.1086/341877},
archivePrefix = {arXiv},
       eprint = {astro-ph/0205257},
 primaryClass = {astro-ph},
       adsurl = {https://ui.adsabs.harvard.edu/abs/2002ApJ...572L.173L}
}

@ARTICLE{2009MNRAS.394..207C,
       author = {{Cao}, Xinwu},
        title = "{An accretion disc-corona model for X-ray spectra of active galactic nuclei}",
      journal = {\mnras},
         year = 2009,
        month = mar,
       volume = {394},
       number = {1},
        pages = {207-213},
          doi = {10.1111/j.1365-2966.2008.14347.x},
archivePrefix = {arXiv},
       eprint = {0812.1828},
 primaryClass = {astro-ph},
       adsurl = {https://ui.adsabs.harvard.edu/abs/2009MNRAS.394..207C}
}

@ARTICLE{2012ApJ...761..109Y,
       author = {{You}, Bei and {Cao}, Xinwu and {Yuan}, Ye-Fei},
        title = "{A General Relativistic Model of Accretion Disks with Coronae Surrounding Kerr Black Holes}",
      journal = {\apj},
         year = 2012,
        month = dec,
       volume = {761},
       number = {2},
          eid = {109},
        pages = {109},
          doi = {10.1088/0004-637X/761/2/109},
archivePrefix = {arXiv},
       eprint = {1210.2662},
 primaryClass = {astro-ph.HE},
       adsurl = {https://ui.adsabs.harvard.edu/abs/2012ApJ...761..109Y}
}

@ARTICLE{1992MNRAS.259..604T,
       author = {{Tout}, C.~A. and {Pringle}, J.~E.},
        title = "{Accretion disc viscosity: a simple model for a magnetic dynamo.}",
      journal = {\mnras},
         year = 1992,
        month = dec,
       volume = {259},
        pages = {604-612},
          doi = {10.1093/mnras/259.4.604},
       adsurl = {https://ui.adsabs.harvard.edu/abs/1992MNRAS.259..604T}
}

@ARTICLE{1998MNRAS.299L..15D,
       author = {{Di Matteo}, T.},
        title = "{Magnetic reconnection: flares and coronal heating in active galactic nuclei}",
      journal = {\mnras},
         year = 1998,
        month = aug,
       volume = {299},
       number = {1},
        pages = {L15-l20},
          doi = {10.1046/j.1365-8711.1998.01950.x},
archivePrefix = {arXiv},
       eprint = {astro-ph/9805347},
 primaryClass = {astro-ph},
       adsurl = {https://ui.adsabs.harvard.edu/abs/1998MNRAS.299L..15D}
}

@ARTICLE{2000ApJ...534..398M,
       author = {{Miller}, Kristen A. and {Stone}, James M.},
        title = "{The Formation and Structure of a Strongly Magnetized Corona above a Weakly Magnetized Accretion Disk}",
      journal = {\apj},
         year = 2000,
        month = may,
       volume = {534},
       number = {1},
        pages = {398-419},
          doi = {10.1086/308736},
archivePrefix = {arXiv},
       eprint = {astro-ph/9912135},
 primaryClass = {astro-ph},
       adsurl = {https://ui.adsabs.harvard.edu/abs/2000ApJ...534..398M}
}

@ARTICLE{2016ApJ...833...35L,
       author = {{Liu}, J.~Y. and {Qiao}, E.~L. and {Liu}, B.~F.},
        title = "{Revisiting the Structure and Spectrum of the Magnetic-reconnection-heated Corona in Luminous AGNs}",
      journal = {\apj},
         year = 2016,
        month = dec,
       volume = {833},
       number = {1},
          eid = {35},
        pages = {35},
          doi = {10.3847/1538-4357/833/1/35},
archivePrefix = {arXiv},
       eprint = {1611.07149},
 primaryClass = {astro-ph.HE},
       adsurl = {https://ui.adsabs.harvard.edu/abs/2016ApJ...833...35L}
}

@ARTICLE{2020MNRAS.495.1158C,
       author = {{Cheng}, Huaqing and {Liu}, B.~F. and {Liu}, Jieying and {Liu}, Zhu and {Qiao}, Erlin and {Yuan}, Weimin},
        title = "{Magnetic-reconnection-heated corona in active galactic nuclei: refined disc-corona model and application to broad-band radiation}",
      journal = {\mnras},
         year = 2020,
        month = jun,
       volume = {495},
       number = {1},
        pages = {1158-1171},
          doi = {10.1093/mnras/staa1250},
archivePrefix = {arXiv},
       eprint = {2006.02665},
 primaryClass = {astro-ph.HE},
       adsurl = {https://ui.adsabs.harvard.edu/abs/2020MNRAS.495.1158C}
}

@ARTICLE{1984PASJ...36...71M,
       author = {{Matsumoto}, R. and {Kato}, S. and {Fukue}, J. and {Okazaki}, A.~T.},
        title = "{Viscous transonic flow around the inner edge of geometrically thin accretion disks}",
      journal = {\pasj},
         year = 1984,
        month = jan,
       volume = {36},
       number = {1},
        pages = {71-85},
       adsurl = {https://ui.adsabs.harvard.edu/abs/1984PASJ...36...71M}
}

@ARTICLE{2000PASJ...52..133W,
       author = {{Watarai}, Ken-ya and {Fukue}, Jun and {Takeuchi}, Mitsuru and {Mineshige}, Shin},
        title = "{Galactic Black-Hole Candidates Shining at the Eddington Luminosity}",
      journal = {\pasj},
         year = 2000,
        month = feb,
       volume = {52},
        pages = {133},
          doi = {10.1093/pasj/52.1.133},
       adsurl = {https://ui.adsabs.harvard.edu/abs/2000PASJ...52..133W}
}

@ARTICLE{2024ApJ...969..104F,
       author = {{Faris}, Sara and {Arcavi}, Iair and {Makrygianni}, Lydia and {Hiramatsu}, Daichi and {Terreran}, Giacomo and {Farah}, Joseph and {Howell}, D. Andrew and {McCully}, Curtis and {Newsome}, Megan and {Padilla Gonzalez}, Estefania and {Pellegrino}, Craig and {Bostroem}, K. Azalee and {Abojanb}, Wiam and {Lam}, Marco C. and {Tomasella}, Lina and {Brink}, Thomas G. and {Filippenko}, Alexei V. and {French}, K. Decker and {Clark}, Peter and {Graur}, Or and {Leloudas}, Giorgos and {Gromadzki}, Mariusz and {Anderson}, Joseph P. and {Nicholl}, Matt and {Guti{\'e}rrez}, Claudia P. and {Kankare}, Erkki and {Inserra}, Cosimo and {Galbany}, Llu{\'\i}s and {Reynolds}, Thomas and {Mattila}, Seppo and {Heikkil{\"a}}, Teppo and {Wang}, Yanan and {Onori}, Francesca and {Wevers}, Thomas and {Coughlin}, Eric R. and {Charalampopoulos}, Panos and {Johansson}, Joel},
        title = "{Light-curve Structure and H{\ensuremath{\alpha}} Line Formation in the Tidal Disruption Event AT 2019azh}",
      journal = {\apj},
         year = 2024,
        month = jul,
       volume = {969},
       number = {2},
          eid = {104},
        pages = {104},
          doi = {10.3847/1538-4357/ad4a72},
archivePrefix = {arXiv},
       eprint = {2312.03842},
 primaryClass = {astro-ph.HE},
       adsurl = {https://ui.adsabs.harvard.edu/abs/2024ApJ...969..104F}
}

@ARTICLE{2024ApJ...966..160G,
       author = {{Guolo}, Muryel and {Gezari}, Suvi and {Yao}, Yuhan and {van Velzen}, Sjoert and {Hammerstein}, Erica and {Cenko}, S. Bradley and {Tokayer}, Yarone M.},
        title = "{A Systematic Analysis of the X-Ray Emission in Optically Selected Tidal Disruption Events: Observational Evidence for the Unification of the Optically and X-Ray-selected Populations}",
      journal = {\apj},
         year = 2024,
        month = may,
       volume = {966},
       number = {2},
          eid = {160},
        pages = {160},
          doi = {10.3847/1538-4357/ad2f9f},
archivePrefix = {arXiv},
       eprint = {2308.13019},
 primaryClass = {astro-ph.HE},
       adsurl = {https://ui.adsabs.harvard.edu/abs/2024ApJ...966..160G}
}

@ARTICLE{2022ApJ...937....8Y,
       author = {{Yao}, Yuhan and {Lu}, Wenbin and {Guolo}, Muryel and {Pasham}, Dheeraj R. and {Gezari}, Suvi and {Gilfanov}, Marat and {Gendreau}, Keith C. and {Harrison}, Fiona and {Cenko}, S. Bradley and {Kulkarni}, S.~R. and {Miller}, Jon M. and {Walton}, Dominic J. and {Garc{\'\i}a}, Javier A. and {van Velzen}, Sjoert and {Alexander}, Kate D. and {Miller-Jones}, James C.~A. and {Nicholl}, Matt and {Hammerstein}, Erica and {Medvedev}, Pavel and {Stern}, Daniel and {Ravi}, Vikram and {Sunyaev}, R. and {Bloom}, Joshua S. and {Graham}, Matthew J. and {Kool}, Erik C. and {Mahabal}, Ashish A. and {Masci}, Frank J. and {Purdum}, Josiah and {Rusholme}, Ben and {Sharma}, Yashvi and {Smith}, Roger and {Sollerman}, Jesper},
        title = "{The Tidal Disruption Event AT2021ehb: Evidence of Relativistic Disk Reflection, and Rapid Evolution of the Disk-Corona System}",
      journal = {\apj},
         year = 2022,
        month = sep,
       volume = {937},
       number = {1},
          eid = {8},
        pages = {8},
          doi = {10.3847/1538-4357/ac898a},
archivePrefix = {arXiv},
       eprint = {2206.12713},
 primaryClass = {astro-ph.HE},
       adsurl = {https://ui.adsabs.harvard.edu/abs/2022ApJ...937....8Y}
}

@ARTICLE{2000PASJ...52..499M,
       author = {{Mineshige}, Shin and {Kawaguchi}, Toshihiro and {Takeuchi}, Mitsuru and {Hayashida}, Kiyoshi},
        title = "{Slim-Disk Model for Soft X-Ray Excess and Variability of Narrow-Line Seyfert 1 Galaxies}",
      journal = {\pasj},
         year = 2000,
        month = jun,
       volume = {52},
        pages = {499-508},
          doi = {10.1093/pasj/52.3.499},
archivePrefix = {arXiv},
       eprint = {astro-ph/0003017},
 primaryClass = {astro-ph},
       adsurl = {https://ui.adsabs.harvard.edu/abs/2000PASJ...52..499M}
}

@ARTICLE{1999PASJ...51..725W,
       author = {{Watarai}, Ken-ya and {Fukue}, Jun},
        title = "{Radiative Disk Winds from a Self-Similar Slim Disk}",
      journal = {\pasj},
         year = 1999,
        month = oct,
       volume = {51},
        pages = {725},
          doi = {10.1093/pasj/51.5.725},
       adsurl = {https://ui.adsabs.harvard.edu/abs/1999PASJ...51..725W}
}

@ARTICLE{1999ApJ...516..420W,
       author = {{Wang}, Jian-Min and {Zhou}, You-Yuan},
        title = "{Self-similar Solution of Optically Thick Advection-dominated Flows}",
      journal = {\apj},
         year = 1999,
        month = may,
       volume = {516},
       number = {1},
        pages = {420-424},
          doi = {10.1086/307080},
       adsurl = {https://ui.adsabs.harvard.edu/abs/1999ApJ...516..420W}
}

@ARTICLE{2006ApJ...648..523W,
       author = {{Watarai}, Ken-ya},
        title = "{New Analytical Formulae for Supercritical Accretion Flows}",
      journal = {\apj},
         year = 2006,
        month = sep,
       volume = {648},
       number = {1},
        pages = {523-533},
          doi = {10.1086/505854},
archivePrefix = {arXiv},
       eprint = {astro-ph/0605248},
 primaryClass = {astro-ph},
       adsurl = {https://ui.adsabs.harvard.edu/abs/2006ApJ...648..523W}
}

@ARTICLE{1977SvA....21..708P,
       author = {{Pozdnyakov}, L.~A. and {Sobol}, I.~M. and {Syunyaev}, R.~A.},
        title = "{Effect of the multiple Compton scatterings on an x-ray emission spectrum. Calculation by the Monte Carlo method}",
      journal = {\sovast},
         year = 1977,
        month = dec,
       volume = {21},
        pages = {708-714},
       adsurl = {https://ui.adsabs.harvard.edu/abs/1977SvA....21..708P}
}

@ARTICLE{1983ASPRv...2..189P,
       author = {{Pozdnyakov}, L.~A. and {Sobol}, I.~M. and {Syunyaev}, R.~A.},
        title = "{Comptonization and the shaping of X-ray source spectra - Monte Carlo calculations}",
      journal = {\apspr},
         year = 1983,
        month = jan,
       volume = {2},
        pages = {189-331},
       adsurl = {https://ui.adsabs.harvard.edu/abs/1983ASPRv...2..189P}
}

@ARTICLE{2021MNRAS.500.1673H,
       author = {{Hinkle}, Jason T. and {Holoien}, T.~W. -S. and {Auchettl}, K. and {Shappee}, B.~J. and {Neustadt}, J.~M.~M. and {Payne}, A.~V. and {Brown}, J.~S. and {Kochanek}, C.~S. and {Stanek}, K.~Z. and {Graham}, M.~J. and {Tucker}, M.~A. and {Do}, A. and {Anderson}, J.~P. and {Bose}, S. and {Chen}, P. and {Coulter}, D.~A. and {Dimitriadis}, G. and {Dong}, Subo and {Foley}, R.~J. and {Huber}, M.~E. and {Hung}, T. and {Kilpatrick}, C.~D. and {Pignata}, G. and {Piro}, A.~L. and {Rojas-Bravo}, C. and {Siebert}, M.~R. and {Stalder}, B. and {Thompson}, Todd A. and {Tonry}, J.~L. and {Vallely}, P.~J. and {Wisniewski}, J.~P.},
        title = "{Discovery and follow-up of ASASSN-19dj: an X-ray and UV luminous TDE in an extreme post-starburst galaxy}",
      journal = {\mnras},
         year = 2021,
        month = jan,
       volume = {500},
       number = {2},
        pages = {1673-1696},
          doi = {10.1093/mnras/staa3170},
archivePrefix = {arXiv},
       eprint = {2006.06690},
 primaryClass = {astro-ph.HE},
       adsurl = {https://ui.adsabs.harvard.edu/abs/2021MNRAS.500.1673H}
}

@ARTICLE{2002RvMA...15...27K,
       author = {{Komossa}, Stefanie},
        title = "{Ludwig Biermann Award Lecture: X-ray Evidence for Supermassive Black Holes at the Centers of Nearby, Non-Active Galaxies}",
      journal = {Reviews in Modern Astronomy},
         year = 2002,
        month = jan,
       volume = {15},
        pages = {27},
          doi = {10.48550/arXiv.astro-ph/0209007},
archivePrefix = {arXiv},
       eprint = {astro-ph/0209007},
 primaryClass = {astro-ph},
       adsurl = {https://ui.adsabs.harvard.edu/abs/2002RvMA...15...27K}
}

@ARTICLE{2015JHEAp...7..148K,
       author = {{Komossa}, S.},
        title = "{Tidal disruption of stars by supermassive black holes: Status of observations}",
      journal = {Journal of High Energy Astrophysics},
         year = 2015,
        month = sep,
       volume = {7},
        pages = {148-157},
          doi = {10.1016/j.jheap.2015.04.006},
archivePrefix = {arXiv},
       eprint = {1505.01093},
 primaryClass = {astro-ph.HE},
       adsurl = {https://ui.adsabs.harvard.edu/abs/2015JHEAp...7..148K}
}

@ARTICLE{1990Sci...247..817R,
       author = {{Rees}, Martin J.},
        title = "{``Dead Quasars'' in Nearby Galaxies?}",
      journal = {Science},
         year = 1990,
        month = feb,
       volume = {247},
       number = {4944},
        pages = {817-823},
          doi = {10.1126/science.247.4944.817},
       adsurl = {https://ui.adsabs.harvard.edu/abs/1990Sci...247..817R}
}

@ARTICLE{2017MNRAS.467..898Q,
       author = {{Qiao}, Erlin and {Liu}, B.~F.},
        title = "{The condensation of the corona for the correlation between the hard X-ray photon index {\ensuremath{\Gamma}} and the reflection scaling factor ℜ in active galactic nuclei}",
      journal = {\mnras},
         year = 2017,
        month = may,
       volume = {467},
       number = {1},
        pages = {898-905},
          doi = {10.1093/mnras/stx121},
archivePrefix = {arXiv},
       eprint = {1701.04211},
 primaryClass = {astro-ph.HE},
       adsurl = {https://ui.adsabs.harvard.edu/abs/2017MNRAS.467..898Q}
}

@ARTICLE{2018MNRAS.477..210Q,
       author = {{Qiao}, Erlin and {Liu}, B.~F.},
        title = "{A systematic study of the condensation of the corona and the application for {\ensuremath{\Gamma}} $_{2-10 keV}$-L$_{bol}$/L$_{Edd}$ correlation in luminous active galactic nuclei}",
      journal = {\mnras},
         year = 2018,
        month = jun,
       volume = {477},
       number = {1},
        pages = {210-218},
          doi = {10.1093/mnras/sty652},
archivePrefix = {arXiv},
       eprint = {1803.03375},
 primaryClass = {astro-ph.HE},
       adsurl = {https://ui.adsabs.harvard.edu/abs/2018MNRAS.477..210Q}
}

@ARTICLE{2013ApJ...777..102Q,
       author = {{Qiao}, Erlin and {Liu}, B.~F. and {Panessa}, Francesca and {Liu}, J.~Y.},
        title = "{The Disk Evaporation Model for the Spectral Features of Low-luminosity Active Galactic Nuclei}",
      journal = {\apj},
         year = 2013,
        month = nov,
       volume = {777},
       number = {2},
          eid = {102},
        pages = {102},
          doi = {10.1088/0004-637X/777/2/102},
archivePrefix = {arXiv},
       eprint = {1309.0090},
 primaryClass = {astro-ph.HE},
       adsurl = {https://ui.adsabs.harvard.edu/abs/2013ApJ...777..102Q}
}

@ARTICLE{2012ApJ...759...65T,
       author = {{Taam}, Ronald E. and {Liu}, B.~F. and {Yuan}, W. and {Qiao}, E.},
        title = "{Disk Corona Interaction: Mechanism for the Disk Truncation and Spectrum Change in Low-luminosity Active Galactic Nuclei}",
      journal = {\apj},
         year = 2012,
        month = nov,
       volume = {759},
       number = {1},
          eid = {65},
        pages = {65},
          doi = {10.1088/0004-637X/759/1/65},
archivePrefix = {arXiv},
       eprint = {1209.4961},
 primaryClass = {astro-ph.HE},
       adsurl = {https://ui.adsabs.harvard.edu/abs/2012ApJ...759...65T}
}

@ARTICLE{2012ApJ...754...81L,
       author = {{Liu}, J.~Y. and {Liu}, B.~F. and {Qiao}, E.~L. and {Mineshige}, S.},
        title = "{The Structure and Spectral Features of a Thin Disk and Evaporation-fed Corona in High-luminosity Active Galactic Nuclei}",
      journal = {\apj},
         year = 2012,
        month = aug,
       volume = {754},
       number = {2},
          eid = {81},
        pages = {81},
          doi = {10.1088/0004-637X/754/2/81},
archivePrefix = {arXiv},
       eprint = {1205.6958},
 primaryClass = {astro-ph.HE},
       adsurl = {https://ui.adsabs.harvard.edu/abs/2012ApJ...754...81L}
}

@ARTICLE{1998PASJ...50..667K,
       author = {{Kubota}, Aya and {Tanaka}, Yasuo and {Makishima}, Kazuo and {Ueda}, Yoshihiro and {Dotani}, Tadayasu and {Inoue}, Hajime and {Yamaoka}, Kazutaka},
        title = "{Evidence for a Black Hole in the X-Ray Transient GRS 1009-45}",
      journal = {\pasj},
         year = 1998,
        month = dec,
       volume = {50},
        pages = {667-673},
          doi = {10.1093/pasj/50.6.667},
       adsurl = {https://ui.adsabs.harvard.edu/abs/1998PASJ...50..667K}
}

@ARTICLE{1986ApJ...308..635M,
       author = {{Makishima}, K. and {Maejima}, Y. and {Mitsuda}, K. and {Bradt}, H.~V. and {Remillard}, R.~A. and {Tuohy}, I.~R. and {Hoshi}, R. and {Nakagawa}, M.},
        title = "{Simultaneous X-Ray and Optical Observations of GX 339-4 in an X-Ray High State}",
      journal = {\apj},
         year = 1986,
        month = sep,
       volume = {308},
        pages = {635},
          doi = {10.1086/164534},
       adsurl = {https://ui.adsabs.harvard.edu/abs/1986ApJ...308..635M}
}

@ARTICLE{1984PASJ...36..741M,
       author = {{Mitsuda}, K. and {Inoue}, H. and {Koyama}, K. and {Makishima}, K. and {Matsuoka}, M. and {Ogawara}, Y. and {Shibazaki}, N. and {Suzuki}, K. and {Tanaka}, Y. and {Hirano}, T.},
        title = "{Energy spectra of low-mass binary X-ray sources observed from Tenma.}",
      journal = {\pasj},
         year = 1984,
        month = jan,
       volume = {36},
        pages = {741-759},
       adsurl = {https://ui.adsabs.harvard.edu/abs/1984PASJ...36..741M}
}

@ARTICLE{2009MNRAS.400.2070S,
       author = {{Strubbe}, Linda E. and {Quataert}, Eliot},
        title = "{Optical flares from the tidal disruption of stars by massive black holes}",
      journal = {\mnras},
         year = 2009,
        month = dec,
       volume = {400},
       number = {4},
        pages = {2070-2084},
          doi = {10.1111/j.1365-2966.2009.15599.x},
archivePrefix = {arXiv},
       eprint = {0905.3735},
 primaryClass = {astro-ph.CO},
       adsurl = {https://ui.adsabs.harvard.edu/abs/2009MNRAS.400.2070S}
}

@ARTICLE{2016ApJ...830..125J,
       author = {{Jiang}, Yan-Fei and {Guillochon}, James and {Loeb}, Abraham},
        title = "{Prompt Radiation and Mass Outflows from the Stream-Stream Collisions of Tidal Disruption Events}",
      journal = {\apj},
         year = 2016,
        month = oct,
       volume = {830},
       number = {2},
          eid = {125},
        pages = {125},
          doi = {10.3847/0004-637X/830/2/125},
archivePrefix = {arXiv},
       eprint = {1603.07733},
 primaryClass = {astro-ph.HE},
       adsurl = {https://ui.adsabs.harvard.edu/abs/2016ApJ...830..125J}
}

@ARTICLE{2014ApJ...788...48S,
       author = {{Shappee}, B.~J. and {Prieto}, J.~L. and {Grupe}, D. and {Kochanek}, C.~S. and {Stanek}, K.~Z. and {De Rosa}, G. and {Mathur}, S. and {Zu}, Y. and {Peterson}, B.~M. and {Pogge}, R.~W. and {Komossa}, S. and {Im}, M. and {Jencson}, J. and {Holoien}, T.~W.-S. and {Basu}, U. and {Beacom}, J.~F. and {Szczygie{\l}}, D.~M. and {Brimacombe}, J. and {Adams}, S. and {Campillay}, A. and {Choi}, C. and {Contreras}, C. and {Dietrich}, M. and {Dubberley}, M. and {Elphick}, M. and {Foale}, S. and {Giustini}, M. and {Gonzalez}, C. and {Hawkins}, E. and {Howell}, D.~A. and {Hsiao}, E.~Y. and {Koss}, M. and {Leighly}, K.~M. and {Morrell}, N. and {Mudd}, D. and {Mullins}, D. and {Nugent}, J.~M. and {Parrent}, J. and {Phillips}, M.~M. and {Pojmanski}, G. and {Rosing}, W. and {Ross}, R. and {Sand}, D. and {Terndrup}, D.~M. and {Valenti}, S. and {Walker}, Z. and {Yoon}, Y.},
        title = "{The Man behind the Curtain: X-Rays Drive the UV through NIR Variability in the 2013 Active Galactic Nucleus Outburst in NGC 2617}",
      journal = {\apj},
         year = 2014,
        month = jun,
       volume = {788},
       number = {1},
          eid = {48},
        pages = {48},
          doi = {10.1088/0004-637X/788/1/48},
archivePrefix = {arXiv},
       eprint = {1310.2241},
 primaryClass = {astro-ph.HE},
       adsurl = {https://ui.adsabs.harvard.edu/abs/2014ApJ...788...48S}
}

@ARTICLE{2025MNRAS.539.3473Q,
       author = {{Qiao}, Erlin and {Wu}, Yongxin and {Lin}, Yiyang and {Guo}, Meng and {Liu}, Jifeng and {Guo}, Chenlei and {Jin}, Chichuan and {Jiang}, Ning},
        title = "{Early evolution of super-Eddington accretion flow in tidal disruption events}",
      journal = {\mnras},
         year = 2025,
        month = jun,
       volume = {539},
       number = {4},
        pages = {3473-3488},
          doi = {10.1093/mnras/staf719},
archivePrefix = {arXiv},
       eprint = {2505.02434},
 primaryClass = {astro-ph.HE},
       adsurl = {https://ui.adsabs.harvard.edu/abs/2025MNRAS.539.3473Q}
}

@ARTICLE{2025arXiv250916544G,
       author = {{Guo}, Chenlei and {Qiao}, Erlin},
        title = "{Light curves of time-dependent accretion disk in tidal disruption events}",
      journal = {arXiv e-prints},
         year = 2025,
        month = sep,
          eid = {arXiv:2509.16544},
        pages = {arXiv:2509.16544},
          doi = {10.48550/arXiv.2509.16544},
archivePrefix = {arXiv},
       eprint = {2509.16544},
 primaryClass = {astro-ph.HE},
       adsurl = {https://ui.adsabs.harvard.edu/abs/2025arXiv250916544G}
}

@ARTICLE{2013ApJ...764....2Q,
       author = {{Qiao}, Erlin and {Liu}, B.~F.},
        title = "{A Model for the Correlation of Hard X-Ray Index with Eddington Ratio in Black Hole X-Ray Binaries}",
      journal = {\apj},
         year = 2013,
        month = feb,
       volume = {764},
       number = {1},
          eid = {2},
        pages = {2},
          doi = {10.1088/0004-637X/764/1/2},
archivePrefix = {arXiv},
       eprint = {1212.1770},
 primaryClass = {astro-ph.HE},
       adsurl = {https://ui.adsabs.harvard.edu/abs/2013ApJ...764....2Q}
}




\appendix
\section{Self-similar solution of slim disc}\label{sec:A}

\citet{1988ApJ...332..646A} introduced the well-known ``slim disc", which is geometrically and optically thick. In this model, advective energy transport plays a dominant role in the energy balance. The global numerical solutions of the steady-state, transonic disc structure were derived by \citet{1984PASJ...36...71M}, \citet{2000PASJ...52..133W} and \citet{2000PASJ...52..499M}, while self-similar solutions were derived by \citet{1999PASJ...51..725W},\citet{1999ApJ...516..420W} and \citet{2006ApJ...648..523W}. To some extent, the self-similar solutions can still adequately capture the overall characteristics of the slim disc. Therefore, we adopt \citet{2006ApJ...648..523W}'s self-similar solution and assume that the disc is isothermal rather than polytropic in the vertical direction. The hydrodynamic equations are expressed as follows,
\begin{equation}
\label{Equ:equA1}
\dot M = -2\pi R v_{R} \Sigma
\end{equation}
\begin{equation}
v_{R} \frac{dv_{R}}{dR} + \frac{1}{\Sigma} \frac{d\Pi}{dR} = R(\Omega^{2}-\Omega_{K}^{2}) - \frac{\Pi}{\Sigma} \frac{d\ln\Omega_{K}}{dR}
\end{equation}
\begin{equation}
\dot M (l-l_{\text{in}})= -2\pi R^{2} T_{R\varphi}
\end{equation}
\begin{equation}
\frac{\Pi}{\Sigma}=H^{2}\Omega_{K}^{2}
\end{equation}
\begin{equation}
Q_{\text{vis}}^{+}=Q_{\text{adv}}^{-}+Q_{\text{rad}}^{-}
\end{equation}
where $\Sigma$ is the surface density, $\Pi$ is the pressure integrated in the vertical direction, $l$ is the specific angular momentum, $l_{\text{in}}$ is the angular momentum at the inner boundary and $H$ is the half-thickness of the disc. $T_{R\varphi}$  is the $R-\varphi$ component of the viscous stress tensor, we adopted $T_{R\varphi}=-\alpha \Pi$. The viscous heating rate, the advection cooling rate, and the radiation cooling rate are expressed as follows,
\begin{equation}
Q_{\text{vis}}^{+}=RT_{R\varphi}\frac{d\Omega}{dR}
\end{equation}
\begin{equation}
Q_{\text{adv}}^{-}=\frac{\dot M}{2\pi R^{2}} \frac{\Pi}{\Sigma} \xi
\end{equation}
\begin{equation}
Q_{\text{rad}}^{-}=\frac{8acT_{\rm d}^{4}}{3\kappa \rho_{\rm d}H}\approx\frac{8c\Pi}{\kappa_{\text{es}}\Sigma H}
\end{equation}
where $\xi$ is a dimensionless quantity and we set $\xi=1.5$ as \citet{2006ApJ...648..523W}, $\kappa_{\text{es}}$ is the electron scattering opacity, $\kappa \approx \kappa_{\text{es}}$. \citet{2006ApJ...648..523W} also introduced the ratio of the advection cooling rate to the viscous heating rate,
\begin{equation}
\label{Equ:equA9}
f_{\text{a}}=\frac{Q_{\text{adv}}^{-}}{Q_{\text{vis}}^{+}}=\frac{Q_{\text{adv}}^{-}}{Q_{\text{adv}}^{-}+Q_{\text{rad}}^{-}}
\end{equation}
and derived the analytical expression for $f_{\text{a}}$,
\begin{equation}
\label{Equ:equA10}
f_{\text{a}}(\dot m, \hat r)=f_{\text{a}}(x)=0.5(D^{2}x^{2}+2-Dx\sqrt{D^{2}x^{2}+4}) 
\end{equation}
where $\dot m = \dot M /\dot M_{\text{Edd}}$, $\hat r = R / R_{\text{S}}$ ($R_{\text{S}} = 2GM/c^{2}$), $x = \hat r / \dot m $, $D \approx 0.654\phi^{-1/2}$. 
In our numerical calculations, since a fraction $f_{\rm c}$ of the accretion energy is dissipated in the corona, 
$\dot{m}$ should be replaced by $(1-f_{\rm c})\dot{m}$ \citep{2002ApJ...572L.173L}. 
Strictly speaking, $(1-f_{\rm c})\dot{m}$ should be substituted directly into Equation~(\ref{Equ:equA10}), 
however, this leads to a much more complicated expression and we can not obtain an analytical expression for $f_{\rm c}$. For simplicity, we expand $f_{\rm a}$ in Equation~(\ref{Equ:equA10}) and obtain an approximate relation $f_{\rm a} \approx D^{-2} x^{-2} \propto \dot{m}^{2}$. Therefore, we replace $f_{\rm a}$ by $(1-f_{\rm c})^{2} f_{\rm a}$ in the practical numerical computation.

Assuming that the plasma in the disc is composed of pure ionized hydrogen, by using $\Sigma = 2\rho_{\rm d} H = 2 m_{\text{H}} n_{\rm d} H$, and combining equations (\ref{Equ:equA1}) to (\ref{Equ:equA10}), we can derive the disc temperature and number density as follows, 
\begin{equation}
\label{Equ:equA11}
T_{\rm d} \approx 7.37 \times 10^{5}f_{\text{a}}^{-1/8}\alpha_{0.1}^{-1/4}m_{6}^{-1/4}\dot m^{1/4}\hat{r}_{10}^{-5/8}\phi^{1/8}\text{K}
\end{equation}
\begin{equation}
\label{Equ:equA12}
n_{\rm d} \approx 6.64 \times 10^{12}f_{\text{a}}^{-3/2}\alpha_{0.1}^{-1}m_{6}^{-1}\dot m\hat{r}_{10}^{-3/2}\phi^{-1/2}\text{cm}^{-3}.
\end{equation}
Since the radiative flux at each radius is,
\begin{equation}
F = \frac{1}{2} Q_{\text{rad}}^{-} = \frac{16\sigma T_{\rm d}^{4}}{3\tau} = \sigma T_{\text{eff}}^{4}
\end{equation}
and the optical depth $\tau$ was defined by $\tau = \kappa_{\text{es}} \Sigma / 2$, we can obtain the effective temperature distribution,
\begin{equation}
\label{Equ:equA14}
T_{\text{eff}} \approx 5.69 \times 10^{5}f_{\text{a}}^{1/8}m_{6}^{-1/4}\hat{r}_{10}^{-1/2}\phi^{1/8}\text{K}.
\end{equation}


\bsp	
\label{lastpage}
\end{document}